\documentclass[12pt,preprint]{aastex}

\shorttitle{Transport through Inner Disk}
\usepackage{graphicx}
\usepackage{psfig}

\begin{document}

\title{Magnetically Driven Accretion Flows in the Kerr Metric IV:
Dynamical Properties of the Inner Disk}

\author{Julian H. Krolik}
\affil{Physics and Astronomy Department\\
Johns Hopkins University\\
Baltimore, MD 21218}
\and
\author{John F. Hawley}
\affil{Astronomy Department\\
University of Virginia\\
P.O. Box 3818, University Station\\
Charlottesville, VA 22903-0818}
\and
\author{Shigenobu Hirose}
\affil{Physics and Astronomy Department\\
Johns Hopkins University\\
Baltimore, MD 21218}

\email{jhk@pha.jhu.edu; jh8h@virginia.edu; shirose@pha.jhu.edu }

\begin{abstract}

This paper continues the analysis of a set of general relativistic 3D
MHD simulations of accreting tori in the Kerr metric with different
black hole spins.  We focus on bound matter inside the initial pressure
maximum, where the time-averaged motion of gas is inward and an
accretion disk forms.  We use the flows of mass, angular momentum, and
energy in order to understand dynamics in this region.  The sharp
reduction in accretion rate with increasing black hole spin reported in
Paper I of this series is explained by a strongly spin-dependent
outward flux of angular momentum conveyed electromagnetically; when
$a/M \geq 0.9$, this flux can be comparable to the inward angular
momentum flux carried by the matter.  In all cases, there is outward
electromagnetic angular momentum flux throughout the flow; in other
words, contrary to the assertions of traditional accretion disk theory,
there is in general no ``stress edge,'' no surface within which the
stress is zero.  The retardation of accretion in the inner disk by
electromagnetic torques also alters the radial distribution of surface
density, an effect that may have consequences for observable properties
such as Compton reflection.  The net accreted angular momentum is
sufficiently depressed by electromagnetic effects that in the most
rapidly-spinning black holes mass growth can lead to spindown.
Spinning black holes also lose energy by Poynting flux; this rate is
also a strongly increasing function of black hole spin, rising to
$\gtrsim 10\%$ of the rest-mass accretion rate at very high spin.  As
the black hole spins faster, the path of the Poynting flux changes from
being predominantly within the accretion disk to predominantly within
the funnel outflow.

\end{abstract}


\section{Introduction}

In this paper, we continue the analysis of a series of simulations of
accretion disks in the Kerr metric introduced in De Villiers, Hawley,
\& Krolik (2003; hereafter Paper I). The emphasis in this paper will be
on the properties of the accretion flow inside the initial pressure
maximum, where the net time-averaged motion is towards the black hole.
This region of the accretion flow, in the terminology introduced in
Paper I, comprises the inner part of the main disk body, the inner
torus, the plunging region, and the corona.  It is arguably the
most important region of all in an accretion disk because it is
here that most of the energy is released that ultimately powers
the observed radiation and outflows.  Our goal is to understand
the peculiarities of dynamics in this region; our principal tool will
be to analyze the flows of three conserved quantities, mass, energy,
and angular momentum.  By so doing, we aim to shed light on the
accretion process, how it is moderated by black hole spin, and how
energy and angular momentum can be extracted from the spinning black
hole.

Accretion onto black holes differs from accretion onto ordinary objects
in several ways, but two stand out: the absence of stable circular
orbits inside of a critical radius, and the presence of frame-dragging
due to black hole spin.  The absence of stable circular orbits between
the radius of marginal stability $r_{ms}$ and the event horizon means
that in this region matter flows in rapidly and the mass surface
density is much smaller than in the main disk body where matter orbits
stably.  We may therefore talk about the disk having an ``inner edge"
in the vicinity of $r_{ms}$.  However, as emphasized by Krolik \&
Hawley (2002), where one places this inner edge depends on precisely
which disk property is under consideration.  For example, two distinct
edges can be defined by dynamical properties:  the turbulence edge,
where a transition takes place in the character of magnetic field
evolution---from turbulence-driven evolution in the disk proper to
evolution by magnetohydrodynamic (MHD) flux-freezing, and the stress
edge, within which the stress in the accreting fluid goes to zero so
that there is no outward transport of angular momentum.  Another two
are more closely related to photon emission properties: the reflection
edge, where production of Compton reflection features ceases, and the
radiation edge, where intrinsic luminosity detected by distant
observers goes to zero due to reduced radiative efficiency or
relativistic effects.  Fully relativistic simulations can explore the
relation between these observable disk edges, and the detailed dynamics
of the accretion flow.  Defining the location and nature of these edges
is central to linking disk dynamical models to disk observables because
these edges determine, among other things, the global radiative
efficiency of accretion.

The other salient feature of relativistic accretion is the presence of
frame-dragging when the black hole spins.  Frame-dragging introduces a
new dynamical effect into the accretion flow, and the spin of the black
hole represents a potential source of energy output beyond that of the
accretion flow itself.  In a previous paper of this series (De Villiers
et al. 2004) we presented evidence from the simulations that black hole
spin energy helps to power axial jets.  Here we focus on the flow of
angular momentum and energy through the disk and corona, the influence
of the black hole spin on the main accretion flow, and how the
character of that flow depends on that spin.  We find that
frame-dragging and its associated effects can have a dramatic impact on
the rate of accretion, the structure of the inner disk, and the
fundamental flow of energy and angular momentum.

This paper is organized as follows:  After a quick summary of the
simulations and the numerical diagnostics we used to study them
(\S\ref{Prelim}), we devote one section each to the flow through the
disk of the three conserved quantities, mass (\S 3), angular momentum
(\S 4), and energy (\S 5).  In each case, we try to highlight how much
goes where, how it is split between matter and electromagnetic
contributions, and how the situation changes with changing black hole
spin.  Section \ref{summary} summarizes our findings and comments on
astrophysical implications of the numerical results.

\section{Preliminaries \label{Prelim}}

\subsection{Simulation definition}

We solve the equations of ideal MHD in the metric of a rotating black
hole.  The specific form of the equations we solve, and the numerical
algorithm incorporated into the GRMHD code are described in detail in
De Villiers \& Hawley (2003).  For reference we
reiterate here the key terms and the definitions of the primary code
variables.

We work in the Kerr metric, expressed in Boyer-Lindquist coordinates,
$(t,r,\theta,\phi)$, for which the line element has the form,
${ds}^2=g_{t t}\,{dt}^2+2\,g_{t \phi}\,{dt}\,{d \phi}+g_{r r}\,{dr}^2 +
g_{\theta \theta}\,{d \theta}^2 +g_{\phi \phi}\,{d \phi}^2$. We use the
metric signature $(-,+,+,+)$.  The determinant of the 4-metric is $g$,
and $\sqrt{-g} = \alpha\,\sqrt{\gamma}$, where $\alpha$ is the lapse
function, $\alpha=1/\sqrt{-g^{tt}}$, and $\gamma$ is the determinant of
the spatial $3$-metric. We follow the usual convention of using Greek
characters to denote full space-time indices and Roman characters for
purely spatial indices.  We use geometrodynamic units where $G = c =
1$; time and distance are in units of the black hole mass, $M$.

The state of the relativistic test fluid at each point in the spacetime
is described by its density, $\rho$, specific internal energy,
$\epsilon$, 4-velocity, $U^\mu$, and isotropic pressure, $P$.  The
relativistic enthalpy is $h=1 + \epsilon + P/\rho$.  The pressure is
related to $\rho$ and $\epsilon$ through the equation of state of an
ideal gas, $P=\rho\,\epsilon\,(\Gamma-1)$, where $\Gamma$ is the
adiabatic exponent.  For these simulations we take $\Gamma=5/3$.  The
magnetic field of the fluid is described by two sets of variables, the
constrained transport magnetic field, $F_{jk}=[ijk]\,{\cal{B}}^i$, and
magnetic field $4$-vector $\sqrt{4\pi}\,b^\mu =
{}^{*}F^{\mu\nu}U_\nu$.  The ideal MHD condition requires $U^\nu
F_{\mu\nu} = 0$.  The magnetic field $b^\mu$ is included in the
definition of the total four momentum, $S_\mu = (\rho\,h\ +
{\|b\|}^2)\,W\,U_\mu$, where $W$ is the Lorentz factor.  We define the
transport velocity as $V^i = U^i/U^t$. 

In Paper I we presented results of a series of high- and
low-resolution simulations, the KD (Keplerian Disk) set of disk
models.  These models have an initial condition consisting of an
isolated gas torus orbiting near the black hole, with a pressure
maximum at $r \approx 25M$, and a slightly sub-Keplerian initial
distribution of angular momentum throughout.  The initial magnetic
field consists of loops of weak poloidal field lying along isodensity
surfaces within the torus.  In this paper the emphasis will be on the
high-resolution models designated KD0, KDI, KDP, and KDE, which differ
in the spin of the black hole around which they orbit, with $a/M = 0$,
0.5, 0.9 and 0.998 respectively.  These models used $192\times
192\times 64$ $(r,\theta,\phi)$ grid zones. The radial grid is set
using a hyperbolic cosine function to maximize the resolution near the
inner boundary, which is at $r_{in}= 2.05\,M$, $1.90\,M$, $1.45\,M$,
and $1.175\,M$ for models KD0, KDI, KDP, and KDE, respectively. The
outer radial boundary is set to $r_{out}=120 M$ in all cases. The
$\theta$-grid ranges over $0.045\, \pi \le \theta \le 0.955\, \pi$,
with an exponential grid spacing function that concentrates zones near
the equator; reflecting boundary conditions are enforced in the
$\theta$-direction.  The $\phi$-grid spans the quarter plane, $0 \le
\phi \le \pi/2$.

\subsection{Simulation diagnostics}

Three-dimensional numerical simulations generate an enormous amount of
data, only a representative sample of which can be examined. Our
analysis is based on a specific set of volume- and shell-averaged
history data, taken every $M$ in time, and complete data snapshots
taken every $80M$ in time.  Although the details of the history
calculations are given in Paper I, we provide here a brief summary to
clarify the calculations of mass, energy, and angular momentum
transport in the inner torus and plunging region.

The normalized shell-average of a quantity ${\cal Q}$ at radius $r$ is 
calculated using
\begin{equation}\label{avgdef}
\langle{\cal Q}\rangle_{\cal A} (r,t) = {1 \over {\cal A}}
      \int{d \theta\,d\phi\, \sqrt{-g}\, {\cal Q}},
\end{equation}
where the bounds of integration range over the full $\theta$ and $\phi$
computational domains and ${\cal A}$ is the area of the shell. The
density-weighted shell-average of a quantity will be denoted
$\langle{\cal Q}\rangle_\rho 
\equiv \langle \rho\,{\cal Q}\rangle/\langle
\rho \rangle$.  The flux ${\cal F}$
of a given quantity through a shell at radius $r$ is computed using a
similar integral, but the result is not normalized to the shell area.
For example, the accretion rate is defined
\begin{equation}
 \dot M (r) = \langle \rho U^r \rangle
      =  -\int \, d\theta \, d\phi \, \sqrt{-g} \,
       \rho U^r 
\label{mdotdefn}
\end{equation}
In addition to the accretion rate (mass flux) we consider
the total energy flux,
\begin{equation}\label{trt}
\langle{T^r}_t\rangle = \langle{\rho\,h\,U^r\,U_{t}}\rangle
+{1 \over 4\,\pi}\,\langle{F^r}_\alpha\,{F_t}^\alpha\rangle
\end{equation}
and the total angular momentum flux,
\begin{equation}\label{trp}
\langle{T^r}_{\phi}\rangle = \langle{\rho\,h\,U^r\,U_{\phi}}\rangle
+{1 \over 4\,\pi}\,\langle{F^r}_\alpha\,{F_\phi}^\alpha\rangle
\end{equation}
The first term in each of these sums represents the contribution to the 
flux from the fluid. The second term represents the contribution due to 
the magnetic field, where the electromagnetic field strength tensor, 
$F^{\mu \nu}$, reduces to combinations of transport velocities and 
constrained transport (CT) variables, as discussed in Paper I. 

We denote the time-average of a shell-averaged quantity (or flux) as
\begin{equation}
\langle\langle{\cal Q}\rangle\rangle(r) = {1 \over t_{max} - t_{min}}
     \int_{t_{min}}^{t_{max}}\,dt\, \langle{\cal Q}\rangle\,(r,t).
\end{equation}
The time-average of an integrated quantity (e.g., $\dot M$) is denoted
by a single bracket.  The bounds of integration are chosen to analyze
the late-time state of the accretion flow after the initial transient
described in Paper I has passed.  Unless otherwise stated, $t_{min} =
2000\,M$, which corresponds to about 2.5 orbits at the initial pressure
maximum, and $t_{max} = 8100\,M$, the end-time of the simulations. Such
time-averages are important in establishing persistent features in
various diagnostics; by their nature, these averages suppress short
timescale variations, so we will also be interested in \textit{rms} 
fluctuations about the time averages.

The time-history data contains the total contribution on a shell, but
does not distinguish between contributions from bound material (i.e., in
the disk and corona) and unbound (in the axial funnel and funnel-wall
jet).  For the purposes of this paper, where we are focused
on the accretion flow, this is an important distinction.  Thus, for
many quantities, we will compute time-averages using the full data
dumps, with integrations restricted to zones where the specific
total energy satisfies $-h\,U_t < 1$, i.e., including only the
contributions from bound material.

\section{Mass Accretion\label{mass}}

\subsection{Inflow equilibrium and fluctuations about equilibrium}

The initial conditions in these simulations consist of an isolated
torus of gas orbiting around the black hole.  A relevant question is
the degree to which the resulting accretion flow from such an initial
state replicates a time-steady accretion disk.  Outside the initial
torus pressure maximum ($r=25\,M$), the gas must move outward as it
accepts angular momentum from matter at smaller radius.  This portion
of the flow certainly cannot represent the behavior of a statistically
time-steady accretion disk fed from large distances.  Inside the
initial pressure maximum, however, there is inflow throughout the
simulation.  Because the inner edge of the initial torus is at $r = 15\,M$,
well outside the marginally stable orbit, a significant amount of time
is required before the flow reaches the black hole and a Keplerian disk
forms.  After that point in time we can investigate whether the
accretion flow has reached an approximate statistical equilibrium.

\begin{figure}
\epsscale{0.5}
\plotone{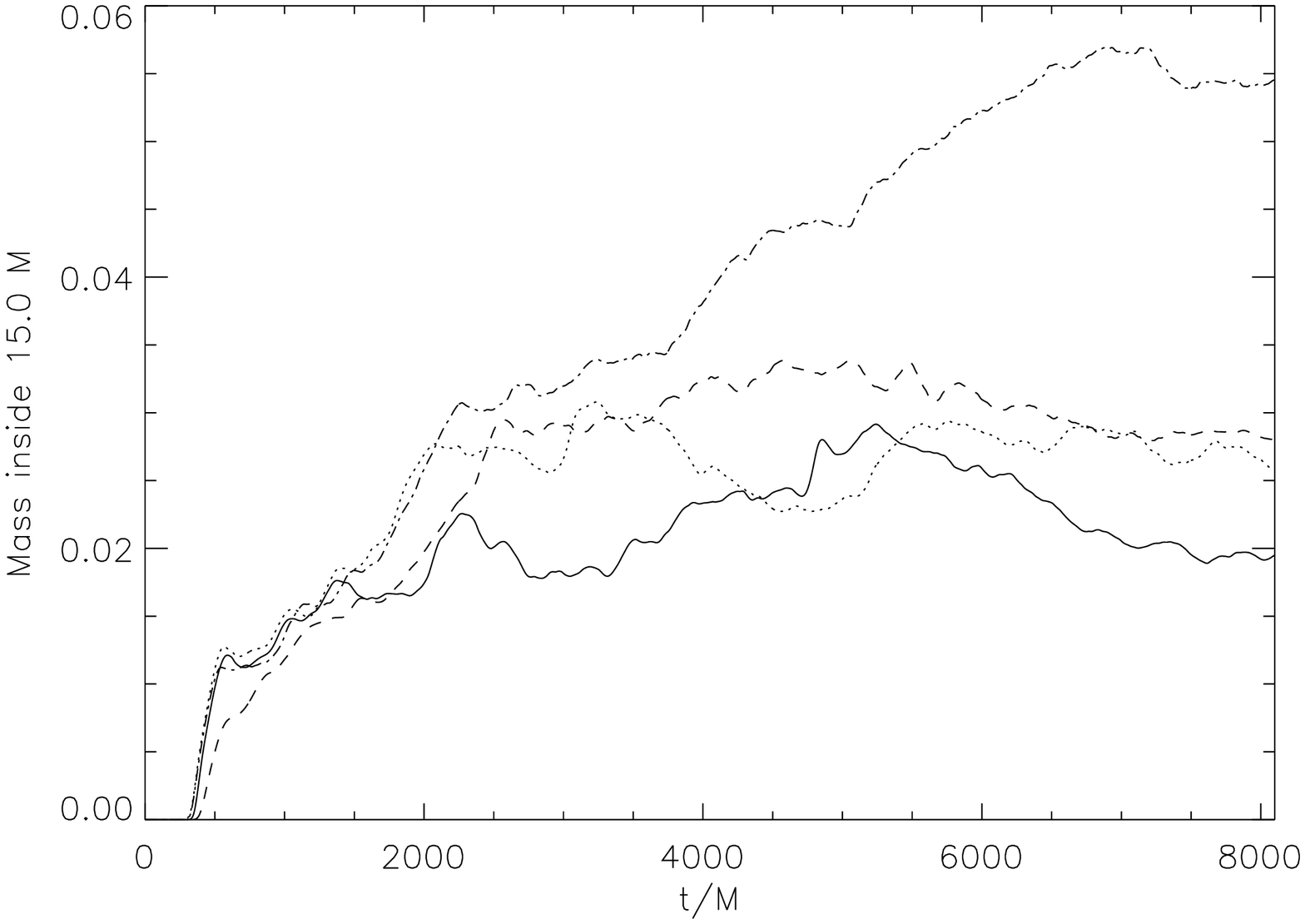}
\caption{\label{Minner}
The total mass found inside $r=15\,M$ as a function of time, given as
fraction of the initial torus mass.
The solid line corresponds to the $a/M = 0.0$ run, the dotted line
to $a/M = 0.5$, the dashed line to $a/M = 0.9$ and the  dot-dashed
line to $a/M = 0.998$.  In the three lower spin models the mass in
the inner accretion disk remains roughly constant after $t/M \approx
2000$, but the mass continues to grow throughout the simulation 
for the high spin model.
}
\end{figure}

We first examine the rate at which gas is fed into the region $r<
15\,M$, namely the region inside of the inner boundary of the initial
torus.  Some sense of the time history can be seen in the spacetime
diagrams of density (Fig.~5 of Paper I) which show both that density
in the inner region is a highly dynamical quantity and that higher
densities are found near the black hole in the high-spin models.
Another view is given in Figure~\ref{Minner}, which presents the total
mass inside of $r=15\,M$ as a function of time for each of the four
models.  In all cases significant matter infall begins by $t=500\,M$
and the mass begins to build up.  After $t = 2000\,M$ in the $a/M = 0$,
0.5, and 0.9 models, the disk in this region appears to settle into a
rough steady state of constant total mass.  The high-spin model
($a/M = 0.998$) stands out: in that model, the mass inside $r=15\,M$
continues to grow throughout most of the simulation.

\begin{figure}
\epsscale{0.75} \plotone{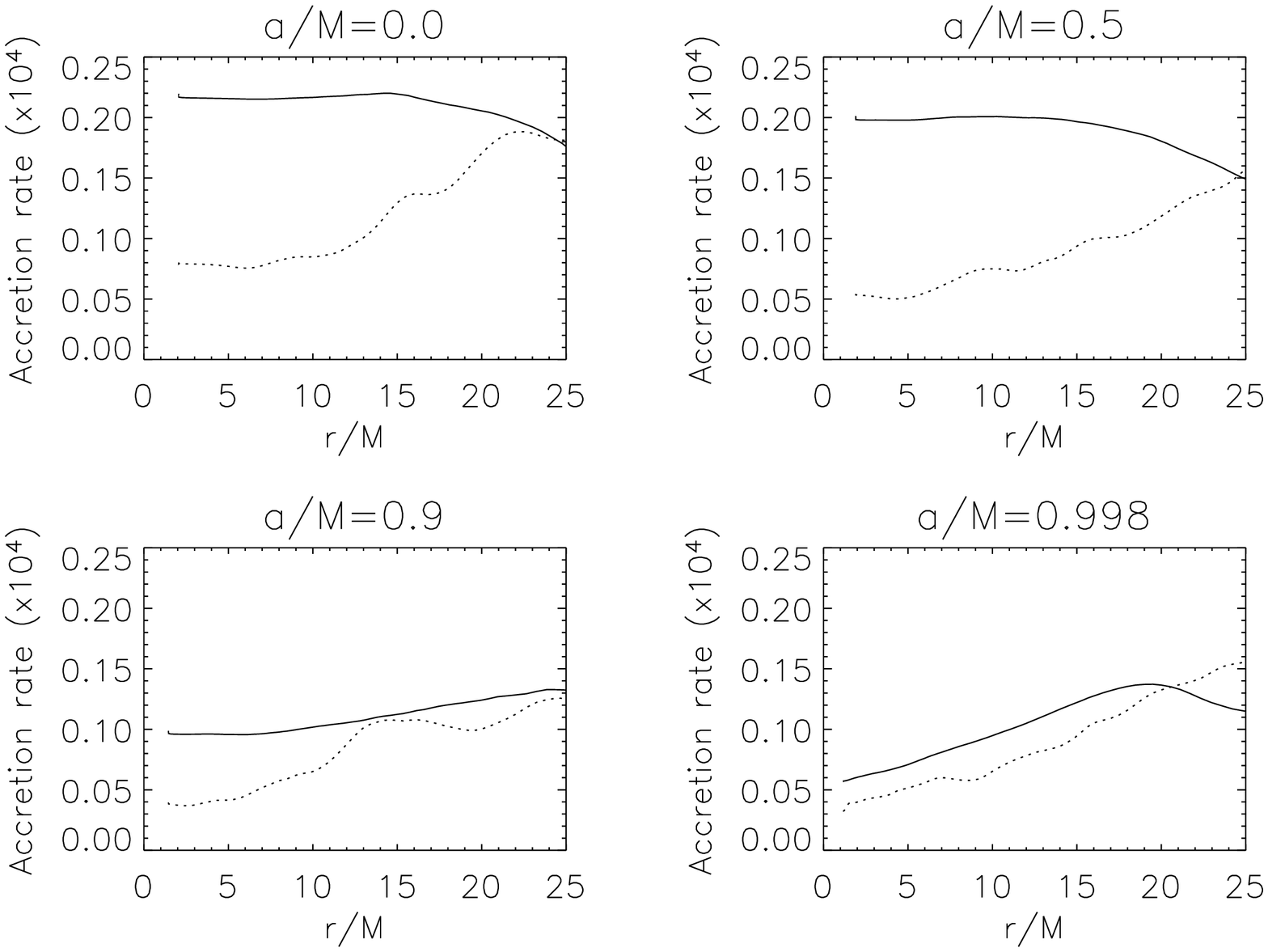}
\caption{\label{mdotfluct}  Time-averaged, shell-integrated accretion
rate as a function of radius (solid curve), and {\it rms} accretion rate
fluctuation about the mean at each radius (dashed curve).  The units of
accretion rate are fraction of the initial torus mass accreted per $M$ of
time (multiplied by $10^4$).
Plots for all four black hole spin models are shown.}
\end{figure}

More detailed information can be obtained by examining the
time-averaged accretion rate as a function of radius
$\langle\dot M (r)\rangle$.  We begin the time average
after $t = 2000\, M$, i.e., after the first 25\% of the total evolution
time.  Figure~\ref{mdotfluct} plots $\langle \dot M (r) \rangle$
for each of the four models.  If the disk is in a statistically steady
state, the time-averaged accretion rate should be nearly constant as a
function of radius.  This is clearly true inside of $r=15\,M$ in the
slowly-rotating simulations ($a/M = 0$, 0.5), and nearly so for the
$a/M =0.9$ model, but is not the case for the $a/M = 0.998$
simulation.  Combining this result with Figure~\ref{Minner}, we
conclude that a state of local accretion equilibrium was attained for
$r \lesssim 10$--$15\,M$ in three of the four simulations.  Although
the $a/M = 0.998$ simulation ran for the same amount of time, 
namely 10 orbits at $r=25M$ and hundreds of orbits in the inner
disk, it did not reach a time-steady state with regard to mass flow.

Despite the constancy with radius of the time-averaged accretion rate,
the \textit{instantaneous} accretion rate varies considerably with
radius and with time.  For example, Fig.~14 of Paper I shows that the
accretion rate into the black hole varies substantially as a function
of time, with significant power across a wide range of timescales.  The
dashed curves in Figure~\ref{mdotfluct} show the {\it rms} deviation
about the local mean accretion rate as a function of radius.  The
relative magnitude of accretion rate fluctuations (as compared to the
mean) increases toward larger radii.  Some of this effect may be
attributed to the initial conditions and the absence of a mean
accretion rate for $r \gtrsim 20M$.  However, a part of this effect is
also likely to be real.  It is significant, for example, that in each
case the fluctuation level begins to rise well within the region where
the accretion rate is nearly constant as a function of radius.

There is also a trend for the relative size of accretion rate
fluctuations at a fixed radius to increase with increasing black hole
spin.  Whereas $\dot M_{rms}/\langle \dot M \rangle\simeq 0.5$ at
$r\simeq 10M$ when $a/M \leq 0.5$, it rises to $\simeq 0.7$ when $a/M
\geq 0.9$; the trend is similar at other radii well inside the initial
pressure maximum ($r=25M$).  Given that the mean accretion rate itself
is poorly-defined when $a/M = 0.998$, it is not surprising that the
{\it rms} fluctuation in the accretion rate in this simulation is 
comparable to the mean at all radii $\lesssim 20M$.

Figure~\ref{mdotfluct} also illustrates a point made in Paper I:  the
fraction of the initial torus mass accreted in a fixed time declines
with increasing black hole spin.  While $14\%$ was accreted over the
duration of the $a/M = 0$ simulation, only $3.5\%$ was accreted when
$a/M = 0.998$.

\subsection{Inflow time}

The mass accretion rate describes the bulk flow of matter toward the
black hole; the inflow time is, on the other hand, a property of
individual particles that make up that matter.  As with many quantities
in fluid dynamics, the inflow time may be studied from either the
Lagrangian or Eulerian points of view.  In this context, the
``Lagrangian" inflow time is the (mean) time for fluid elements to move
from radius $r$ to the inner boundary; the ``Eulerian" inflow time is
the characteristic time for fluid elements to move past a fixed radius
$r$.  The former is of most interest for answering questions such as,
``Does radiation have time to escape before a fluid cell passes the
event horizon?"; the latter is of most interest for gauging the time
for the surface density at some radius to equilibrate with accretion
fluctuations.

\begin{figure}[ht]
\epsscale{0.5}
\centerline{\psfig{file=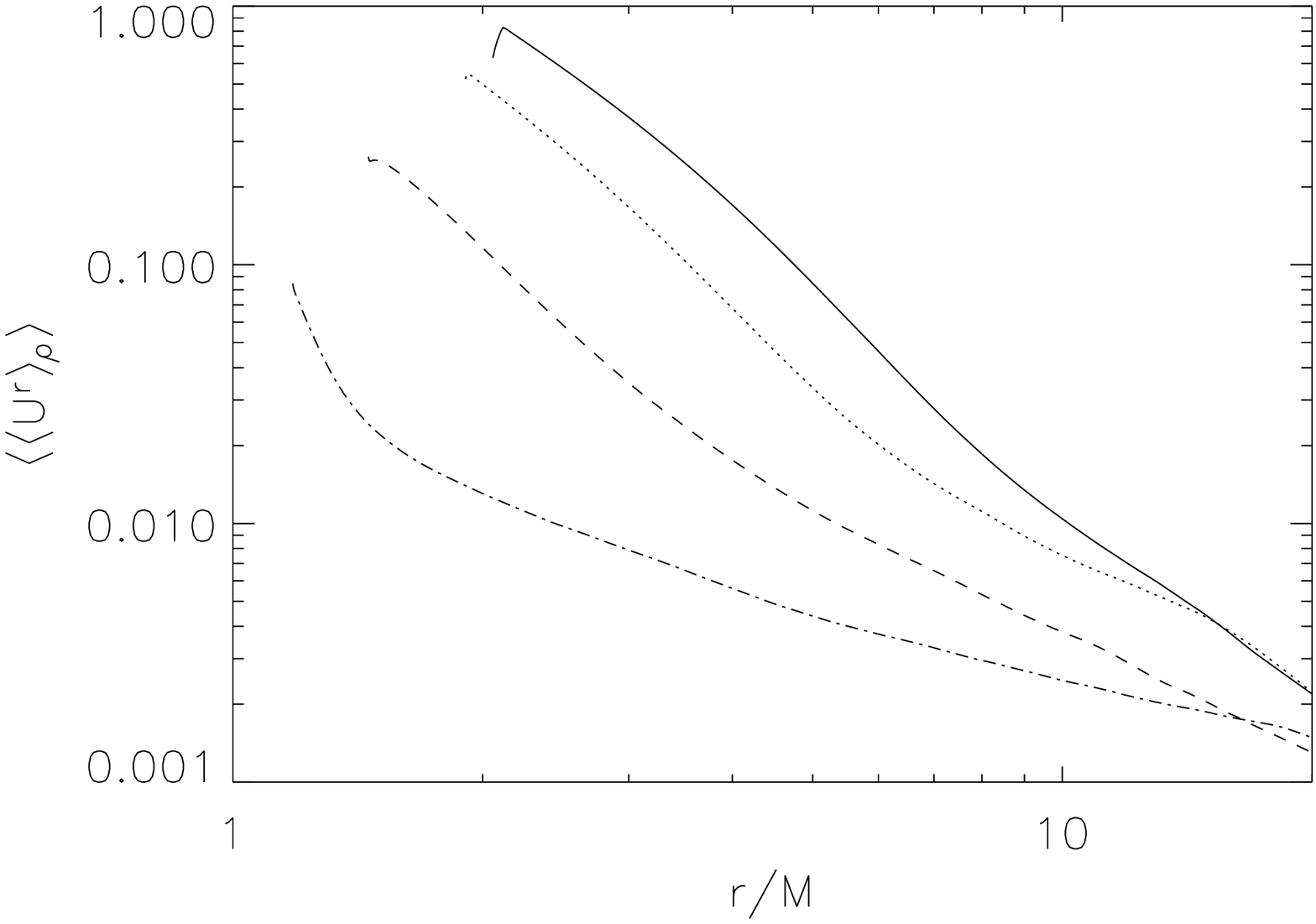,width=2.5in}
       \quad\psfig{file=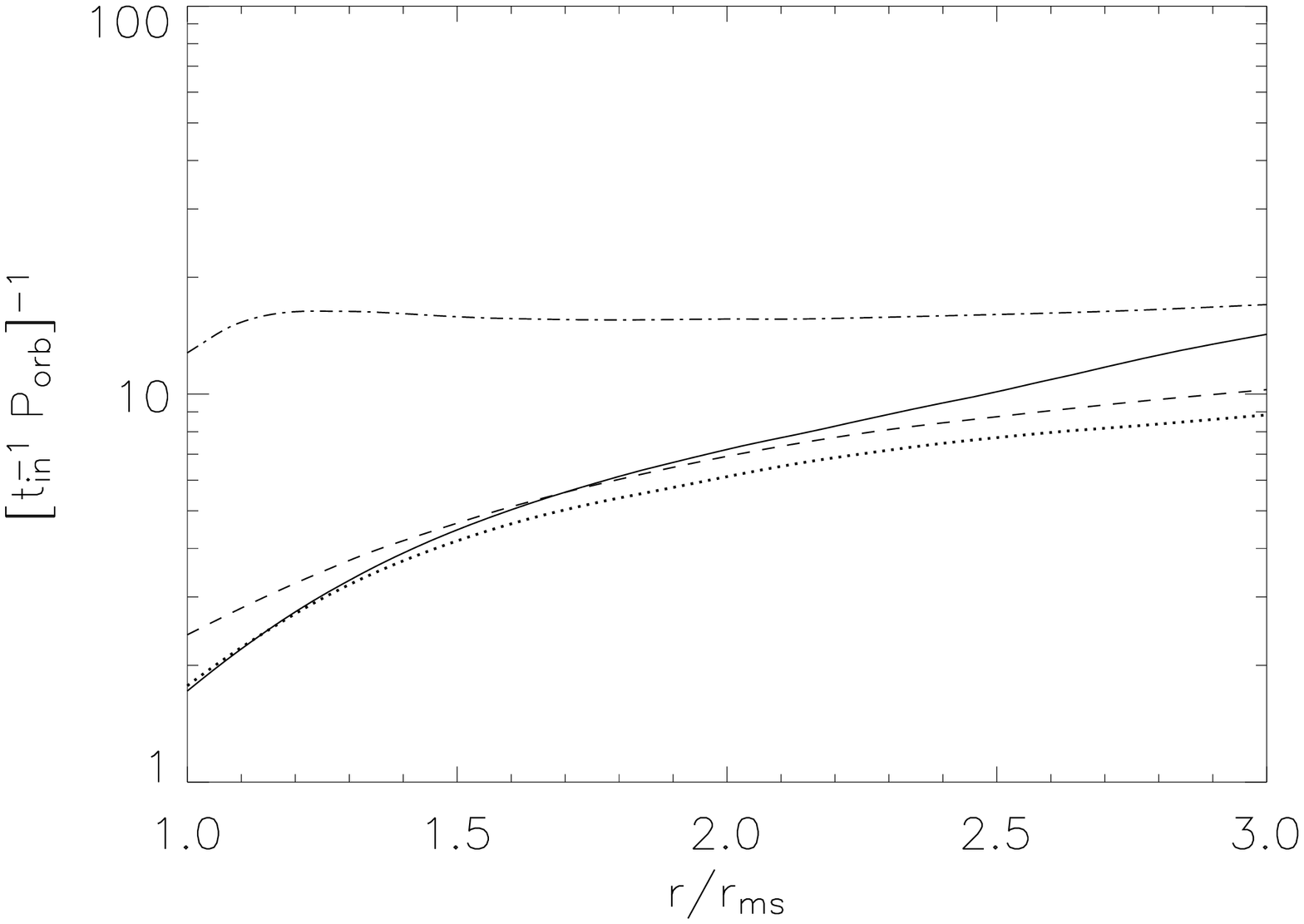,angle=90,width=2.5in}}
\caption{\label{comparetinflow} 
(a) Time-averaged, density-weighted inflow velocity
$\langle\langle U^r \rangle_\rho\rangle$
as a function of $r$. The solid curve is the $a/M
= 0$ simulation, the dotted curve is $a/M = 0.5$, the dashed curve is
$a/M = 0.9$, and the dash-dot curve is $a/M = 0.998$.  (b) Ratio of
inflow time (as defined by eqn.~\ref{inrate}) to orbital period in
the equatorial plane as a function of $r/r_{ms}$.  The curves are
identified as in (a).
}
\end{figure}

We compute both inflow times from the mean inflow velocity obtained from
a time-averaged density-weighted shell-integral.  The quantity
$\langle\langle U^r \rangle_\rho\rangle $ is shown in
Figure~\ref{comparetinflow}a for all
four simulations.  The inflow velocity is smaller with greater
spin $a/M$ at all radii less than about $r=15\,M$.
The slope as a function of $r$ also decreases with increasing 
black hole spin.  Note too that, except for the highest-spin model,
$\langle\langle U^r \rangle_\rho\rangle $ is
very nearly a pure power-law in radius that shows no feature
whatsoever at $r_{ms}$.  

Because it is primarily used for comparisons to equilibration rates,
it is most convenient to define the Eulerian inflow time in terms
of the mean inflow rate,
\begin{equation}\label{inrate}
\langle t_{in}^{-1} \rangle = \langle\langle V^r \rangle_\rho \rangle/r = 
   \left\langle -{\int \, d\theta \, d\phi \sqrt{-g} \rho U^r
\alpha/W\over r \int \, d\theta \, d\phi \sqrt{-g} \rho }
\right\rangle .
\end{equation}
Figure~\ref{comparetinflow}b shows the radial dependence of this inflow
time relative to the orbital period.  In all cases but the most rapidly
spinning black hole, this ratio rises gradually from $\simeq 2$ near
$r=r_{ms}$ to $\simeq 10$--15 at large radii.  The $a/M=0$ and 0.5
models are nearly identical; the $a/M=0.9$ model shows a slightly
flatter curve and a larger value inside $r_{ms}$.  The behavior of this
ratio for the $a/M = 0.998$ model sharply contrasts with that of all
the others: it is 10--20 at all radii.

For a Lagrangian interpretation, one can define the
proper (i.e., fluid frame) radial infall time from a given 
radius to the inner radial boundary by integrating 
\begin{equation} 
\tau_{in} = \int_{r_h}^{r}\,\|ds\| 
\end{equation} 
with $ds^2 = g_{tt}\,dt^2 + g_{rr}\,dr^2$ using the time-averaged
inflow velocity.  Doing so produces curves that are very similar to
those in Figure~\ref{comparetinflow}b, except that the inflow time
is reduced 
by a factor $\sim 3$--$4$.  This  contrast between the Eulerian
and Lagrangian definitions is due to the sharp fall in the ratio of
inflow time to orbital period at smaller radii.  The inflow velocity
scales roughly $\propto r^{-3}$ for the $a/M=0$ case, and is only
slightly less steep when $a/M = 0.5$ or 0.9.  Because of this rapid
increase in velocity near the hole, most of the Lagrangian inflow
time is spent at large radius.

The Lagrangian inflow time provides one estimate for the location of
the turbulence edge, the point where the magnetic field dynamics switch
from being controlled by turbulence to being controlled by flux
freezing in a plunging infall.  Krolik \& Hawley (2002) examined the
ratio of infall to orbital time near the location of the turbulence
edge, and found that it lies near the point where the (Lagrangian)
infall time becomes shorter than the dynamical time.  This occurs at
$\simeq 1.3$--$1.4 r_{ms}$ for all three models with $a/M \leq 0.9$.
When the black hole spins very rapidly, however, this ratio is almost
constant as a function of radius, not betraying a clear location for
the turbulence edge at all.

Indeed, a striking aspect of Figure~\ref{comparetinflow}b is the
contrast between the three more slowly-spinning cases and the most
rapidly-spinning one.  The slow-spin cases behave as one would expect:
well outside the marginally stable orbit, inflow takes multiple orbits
because it is limited by the slow loss of angular momentum due to
magnetic torques.  As one approaches the location of the marginally
stable orbit, however, there is less to prevent inward radial motion at
constant angular momentum because the effective potential is very
nearly flat.  There, the ratio of infall time to orbital period drops
toward one.

The $a/M = 0.998$ case is very different.  This contrast presents a
puzzle because the effective potential seen by its material is not
qualitatively different from the others.  As was discussed in Paper I
(see Fig.~9), the specific angular momentum distribution in the
accretion disk is close to Keplerian in all four simulations.  It is
therefore difficult on the basis of gravitational orbital dynamics
alone to understand why the ratio of inflow to orbital time remains
large even at $r=r_{ms}$ when $a/M = 0.998$, and why the transition
from turbulence to plunging inflow is inhibited.  We believe that this
result has implications for the impact of {\it electromagnetic}
stresses on the angular momentum budget.  This will be discussed in
\S~\ref{ltransport}.

\subsection{Radial distribution of surface density}

\begin{figure}[ht]
\epsscale{0.5}
\centerline{\psfig{file=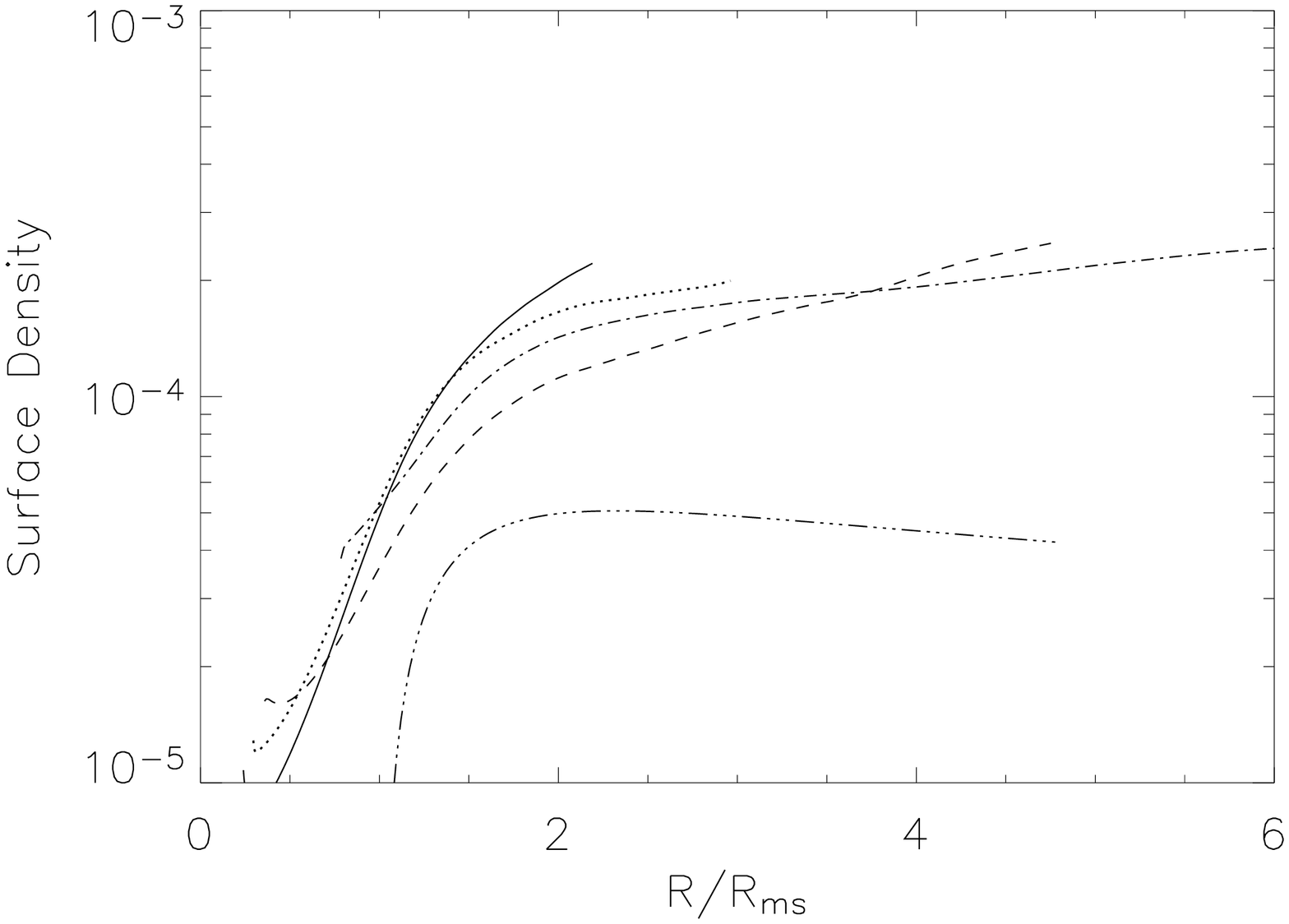,angle=90,width=5.5in}}
\caption{\label{surfdensity} 
Time-averaged surface density $\Sigma$ as a function of proper radius,
$R$, normalized to the marginally stable value, out to
$r=15M$ in each of the four simulations.  Each curve is also normalized
to the initial torus mass of the simulation.  The solid curve is $a/M = 0$,
the dotted is $a/M = 0.5$, the dashed is $a/M = 0.9$, and the
dot-dashed is $a/M = 0.998$.  For comparison, the triple dot-dashed
curve is the surface density predicted by a conventional 
Novikov-Thorne disk model for $a/M = 0.9$,
assuming a constant ratio of stress to pressure.  The amplitude of
the curve is arbitrary; only the shape is relevant.}
\end{figure}

One definition of an accretion disk's inner edge is a rapid decline in
the disk's surface density distribution.  Figure~\ref{surfdensity} is
the time-averaged rest-mass surface density distribution as a function
of radius for all four simulations.  The mean surface density for
each of the simulations is normalized to the initial torus mass for
that simulation.  In addition, in order to avoid creating a misleading
visual impression, the radius in this plot is proper radius, not
Boyer-Lindquist coordinate radius; the relation between this radial variable
and Boyer-Lindquist radius is defined as
\begin{equation}
R = r_{in} + \int_{r_{in}}^{r} \, dr^{\prime} \, \sqrt{g_{rr}(r^{\prime})},
\end{equation}
where the metric coefficient is evaluated in the equatorial plane.
We derive the surface density from
the history data, which provides the total mass in each radial shell for
every $M$ in time.  To obtain a surface density we divide the shell
mass by an equatorial unit area  and average over time for $t >
2000 \, M$.  We restrict radial coverage to regions within $15M$ in order to
consider only those places in approximate inflow equilibrium; for this
reason, when plotted relative to $R_{ms}$, the curves extend for
different distances.  All these curves may be compared with one
illustrating a conventional disk model (triple dot-dashed line).  This curve
was constructed combining the run of stress predicted by the
(zero-stress boundary condition) Novikov-Thorne model and the fixed
ratio of integrated stress to integrated pressure suggested by Shakura
\& Sunyaev (1973).

Several points stand out in Figure~\ref{surfdensity}.  All four
simulations show quite similar behavior, with $\Sigma$ rising
steadily with increasing radius, steeply within the plunging region and
more gradually in the disk proper.  As was emphasized by Krolik \& Hawley
(2002), there is no sharp break in surface density, not even at the
marginally stable radius.  Normalized to the initial mass, the
surface density in the marginally stable region declines by about
a factor of two as $a/M$ increases from 0 to 0.9, but then rises
again when $a/M = 0.998$.  In sharp contrast to this behavior, the
conventional model predicts a well-defined maximum in $\Sigma$
at 2--$3 R_{ms}$, with a drastic drop-off at and inside $R_{ms}$ and
a more gradual decline at larger radius.
The reason for the very sharp break in the standard model is that
the net radial accretion velocity is much smaller than either
the orbital velocity or sound speed outside $R_{ms}$, but is assumed
to abruptly rise to the free-fall velocity inside $R_{ms}$.
The result is an abrupt drop in
surface density.  As can be seen in Figure~\ref{comparetinflow}a, the
accretion velocity in our simulations increases smoothly and continuously
from outside $R_{ms}$ inward toward the hole.

The surface density distribution is important in locating the
reflection edge of the disk, the point within which the disk can no
longer efficiently Compton scatter X-rays (Krolik \& Hawley 2002).
Compton reflection is a process whereby dense, cool matter in the disk
reflects hard X-rays generated somewhere outside the disk, imprinting
upon their spectrum the signature of absorption by heavy element
K-edges.  Efficient reflection requires a disk column density large
enough to make it thick to Compton scattering.  Because the column
density must decrease sharply with diminishing radius in the plunging
region, the reflection edge is generally
assumed to be located at or near $r_{ms}$.

Although the simulations have no absolute density scale, the results
may be phrased in terms of an absolute density once the accretion rate
in Eddington units is specified: the relation is $\Sigma = 38 \, \dot
m$~g~cm$^{-2}$, where the accretion rate relative to Eddington $\dot m$
is defined assuming unit radiative efficiency in rest-mass units ($\dot
M_{Edd} = 1.7 \times 10^{17}\,\dot m M/M_\odot$~g~s${}^{-1}$).  A maximum
value for $\dot m$ in these units would be $\simeq 20$ (if the efficiency
is as low as 0.05 and the luminosity is exactly Eddington), and in most
instances one would expect it be a good deal less.  The surface density
in code units for the three slower-spinning simulations near $r_{ms}$
is $\simeq 0.01$, while it is $\simeq 0.06$ for the $a/M = 0.998$ simulation.
Translating to physical units, we predict Compton optical depths
$\tau_T \simeq 1.3 (\dot m/10)(\Sigma/0.01\hbox{~code units})$.  Thus,
rapidly-accreting systems may indeed find their reflection edges
near $r_{ms}$, but in more slowly-accreting systems it may be significantly
farther out.  And, of course, in all cases, the location of this
edge can vary significantly in time.  Because disk
thermodynamics can influence the radial distribution of surface
density, it may be too early to say definitively how the effects we are
examining alter classical expectations, but it is clear that they can
strongly affect this basic property of disks.

To summarize the results of this section, we find that the low spin
models with $a/M \leq 0.9$ develop accretion disks that are in a
quasi-steady state, but the highest spin model, $a/M=0.998$ does not.
For fixed surface density at large radius, the accretion rate into the
hole decreases with increasing $a/M$, although the accretion rates
feeding mass into the region inside the initial torus are comparable.
This means that the mass of the inner torus tends to be greater for
larger black hole spin.  Accretion infall times, too, are longer with
larger spin.  All these results suggest that the spin of the black hole
affects the disk in ways that go beyond simply determining the location
of the marginally stable orbit.  The mechanism by which the black hole
does so is elucidated by the next topic, the angular momentum fluxes in
the disk.

\section{Angular Momentum Transport}\label{ltransport}

The stress tensor
\begin{equation}
T^\mu_\nu = \rho h U^\mu U_\nu  - b^\mu b_\nu +
||b||^2 U^\mu U_\nu + (p + ||b||^2/2)g^\mu_\nu
\label{stresstensor}
\end{equation}
describes the flux of four-momentum.  Conservation of four-momentum
means that $\nabla_\mu T^\mu_\nu = 0$, where the $\nabla$ indicates
covariant differentiation. The four terms contributing to the stress
tensor each have simple interpretations: the first is the flux directly
associated with matter; the second is the purely electromagnetic part
due to correlations in the magnetic field; the third is the part
proportional to the magnetic energy advected with the flow; the fourth
is the momentum flux associated with pressure.  In the context of
accretion disks, the most interesting component of this tensor is
$T^r_\phi$, the angular momentum flux in the radial direction.
The second term then represents the magnetic torque that moves
angular momentum outward, while the first and third represent the
angular momentum moving inward with the accretion flow, and the fourth
is zero.   Thus, the magnetic torque contribution $-b^r b_\phi$ has
particular importance for accretion disks because it is responsible
for transferring angular momentum outward from one fluid
element to another. The net flux (the sum of all the terms) is the
conserved angular momentum flux through the disk.

\subsection{Shell-integrated angular momentum flux}

\begin{figure}[ht]
\epsscale{0.5}
\centerline{\psfig{figure=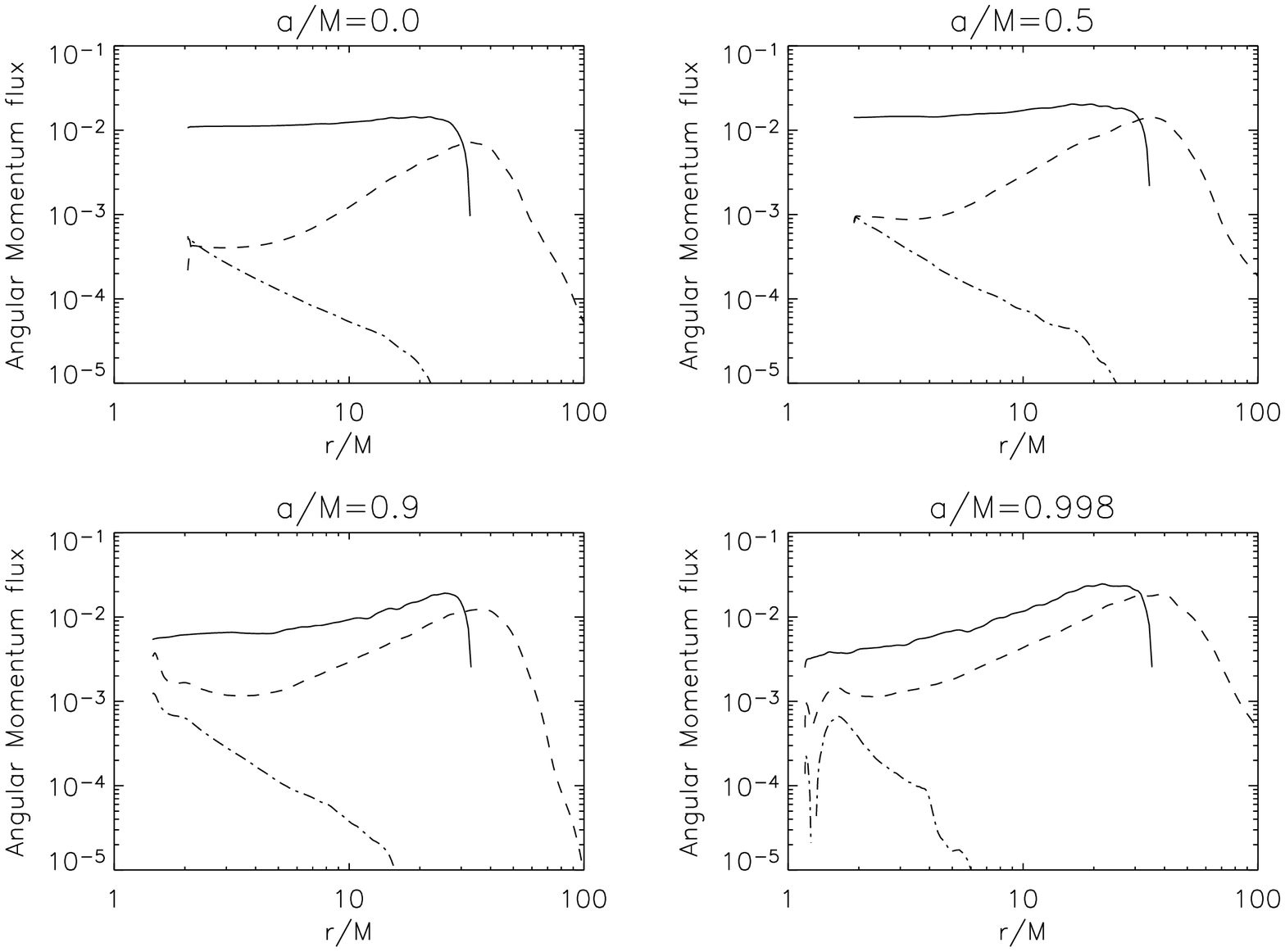,width=5.5in}}
\caption{\label{angmomfluxes} 
Absolute values of the shell-integrated and time-averaged 
angular momentum fluxes in the bound portion of the flow for the four
simulations.  The solid curve is the flux associated with the matter, 
the dashed curve is
the magnetic torque, the dot-dash curve is the part due to advected
magnetic energy.  The sign convention chosen is such that the matter
flux and advected magnetic energy carry prograde angular momentum
inward, while the magnetic torque carries positive angular momentum
outward.}
\end{figure}

To understand better the flow of angular momentum in the disk, it is
helpful to look at the spherical shell-integrated time-average of each
of the three terms in equation~\ref{stresstensor} separately
as a function of radius (Fig.~\ref{angmomfluxes}).  Each of the three
terms is integrated over the angular coordinates, but we restrict the
integration to the region where the flow is bound, i.e., $-hU_t < 1$.
This removes the contributions from the magnetic fields in the funnel
which, as discussed in De Villiers et al. (2004), are not 
insignificant when the black
hole is rotating.  The computation is done using complete data sets
spaced every $80 M$ in time; these are then averaged 
over the last 75\% of the simulation.

Because $\nabla_\mu T^{\mu}_{\nu} = 0$, the
integral over a radial shell, $\int d\Omega \, T^{r}_\nu$, is constant
in a steady-state disk.  Separating the terms of the $r$-$\phi$
component of the stress tensor and employing Gauss's Theorem transforms
this statement into 
\begin{equation}\label{angmomcons} 
\dot M \langle (h + ||b||^2/\rho)U_\phi \rangle - \int \, d\Omega
   \, b^r b_\phi = const.  
\end{equation} 
In conventional disk models it is assumed that
the stress term in equation~(\ref{angmomcons}) is zero inside some radius
that lies well outside the event horizon.  Generally this ``stress
edge" is assumed to lie at the marginally stable orbit, or, in more
detailed hydrodynamical models, it is taken to be the sonic point,
which is generally a short way inside $r_{ms}$ (e.g., 
Abramowicz et al.~1988).  However, as Page \& Thorne (1974) remarked,
and Krolik \& Hawley (2002) discussed in the context of pseudo-Newtonian
models, when magnetic fields are important, the stress edge need not
coincide with $r_{ms}$.  We will shortly examine where it may be found in
our general relativistic simulations.

Time-steadiness also implies that the accretion rate is constant as a
function of radius, so equation~(\ref{angmomcons}) requires the
shell-integrated stress to increase outward to match the angular
momentum per unit mass, which is $\propto r^{1/2}$ in the Newtonian limit.
Physically, this means that for matter to move inward it must lose
angular momentum by transferring the appropriate amount of angular
momentum outward.  A steady state at a fixed position is maintained
when that transfer is balanced by matter flowing in carrying angular
momentum.

We can directly compare this picture to the simulation results.
In Figure~\ref{angmomfluxes} we see that in the low spin cases, $a/M=0.0$
and $a/M=0.5$, some of this picture is reproduced in a time-averaged
sense.  The shell-integrated stress rises outside the marginally stable
orbit, accompanied by a compensating gradient in the matter flux.
However, rather than going to zero inside $r_{ms}$, the shell-integrated
stress, while smaller than in the main disk body, stays at a constant
level.

In the two higher spin models the departures from conventional
expectations are even greater.  Just as in the low-spin cases, stress
continues throughout the plunging region, but its magnitude (there
and in the disk proper) is everywhere proportionally
larger.  In rough terms, its magnitude is comparable to the
angular momentum flux carried with the matter.
We conclude, therefore, that, as viewed in the coordinate frame,
{\it there is no stress edge; stress continues all the way from the
disk body to the event horizon}.

A second, even more dramatic, contrast with conventional
views is that in the inner regions the gradient in the
shell-integrated stress has the wrong sign: the magnetic angular
momentum flow {\it decreases} outward.  Where this is the case,
rotationally-supported matter does not on average lose angular momentum
by magnetic torques.  This
change in sign of the gradient of the angular momentum flow appears to
be the reason why the rate of accretion onto the black hole is smaller
in the two simulations with rapidly-spinning black holes, and likewise
why the mass of the inner torus becomes large in these two cases.

\subsection{Mean specific angular momentum}

\begin{figure}[ht]
\epsscale{0.8}
\plotone{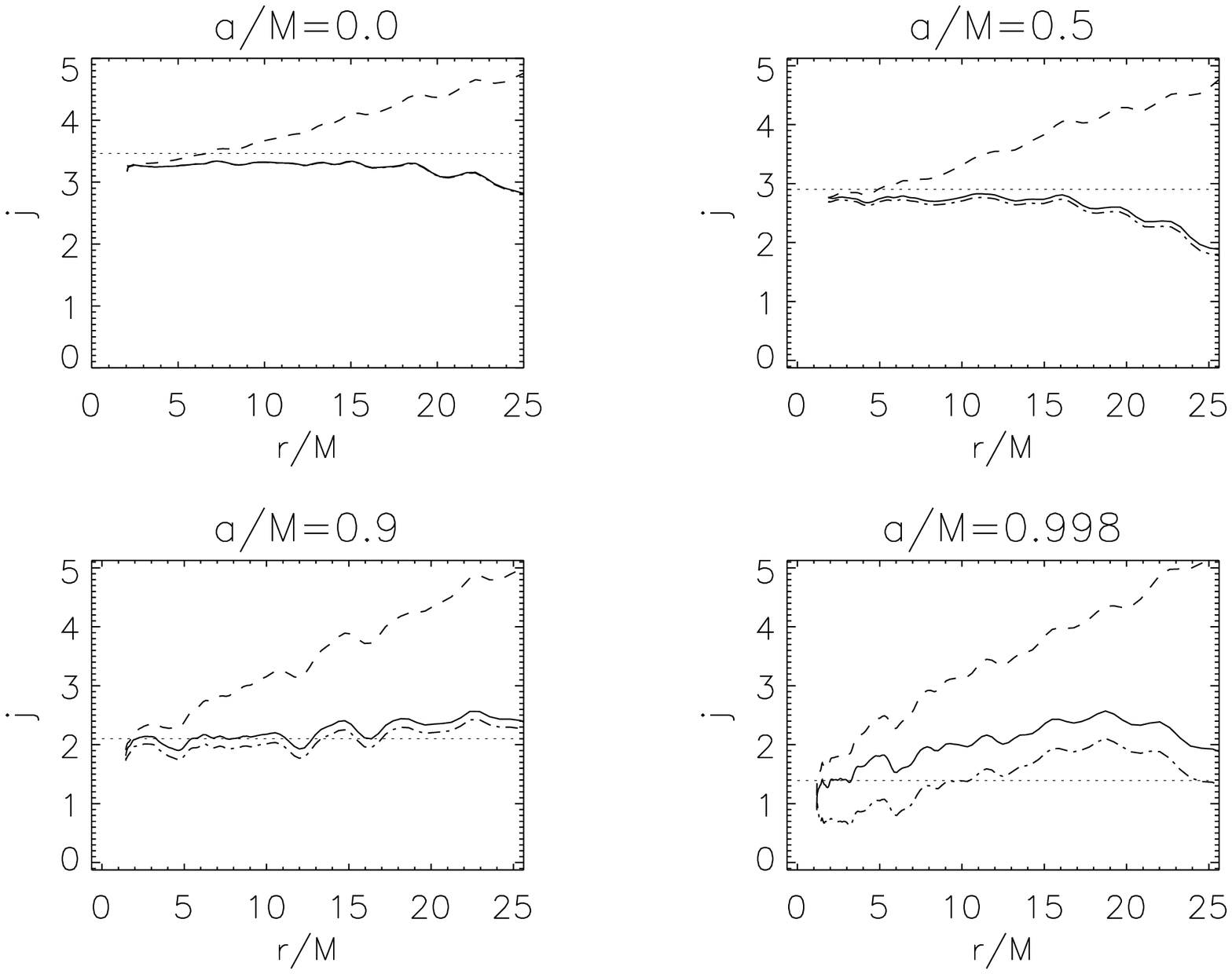}
\caption{\label{specangmom} 
The time-average flux of angular momentum per accreted rest-mass,
$\langle\langle T^r_\phi \rangle\rangle/\langle\dot M\rangle$ 
for each of the four runs.  The data are
time-averaged over the last 75\% of the simulation.  The dashed line is
the angular momentum carried by the matter flux; the solid line is
the total angular momentum flux.  For both these curves the space
integrals are restricted to the bound material, i.e., cells with
$-hU_t < 1$.  The dot-dashed line shows
the total angular momentum flux over the whole domain, including the
jet.  The values are contrasted with the angular momentum per rest-mass
of the marginally stable orbit $j(r_{ms})$ (dotted line).}
\end{figure}

Additional information can be obtained through an analysis of the mean
accreted angular momentum per unit rest-mass, $j$.  Again with the 
integrals restricted to the bound matter, this is defined as
$j(r) = \langle\langle T^r_\phi \rangle\rangle / \langle \dot M\rangle $.
We plot the function $j(r)$ in
Figure~\ref{specangmom}.  Here the solid line shows the specific
angular momentum associated with the total angular momentum flux, while
the dashed line is that of the matter flux alone.  
The $a/M = 0$ and $a/M = 0.5$ simulations behave very much in line with
expectations; $j$ is very close to constant with radius through the
disk, indicating that the angular momentum flux has equilibrated to an
approximately time-steady state throughout these regions.  Although
some angular momentum is carried in the jet when $a/M = 0.5$,
it is too little to have much effect on the total.  In each
model the dotted line indicates the value $j(r_{ms})$ associated with
the marginally stable orbit, and it is clear that $j_{in} < j(r_{ms})$,
where $j_{in}$ is the value carried into the black hole.  This
demonstrates that torques inside the marginally stable orbit do remove
some angular momentum from material flowing through that region, but
the effect is not dramatically large: it ranges from 6\% to 8\%.

In the two simulations with the more rapidly-spinning black holes, the
radial slope in $j(r)$ provides evidence that they are not in steady state,
particularly the  $a/M=0.998$ model.  Again the specific angular
momentum carried by the mass does not differ much from the value
$j(r)$.  However, in these cases there is a significant difference
between the net angular momentum carried by the bound material
and the total.  The angular momentum flux associated with the jet
is $\sim 10\%$ of that carried in bound matter when $a/M = 0.9$, but
is comparable to the flux through the bound matter when $a/M = 0.998$.
When all is said and done,
$j$ is 12\% below $j(r_{ms})$ for the $a/M = 0.9$
model, and 42\% below for $a/M = 0.998$.

   If the conditions of accretion remain constant long enough for
the accreted mass to dominate the original mass in the black hole,
$j$ determines $a/M$.  As noted in Paper I, with $j_{in} = 0.8$
in the $a/M = 0.998$ simulation, in the long run the
spin of the black hole can not be maintained (see also Gammie,
Shapiro, \& McKinney 2004).  

\subsection{Angular distribution of angular momentum flux}

\begin{figure}[ht]
\epsscale{0.8}
\plotone{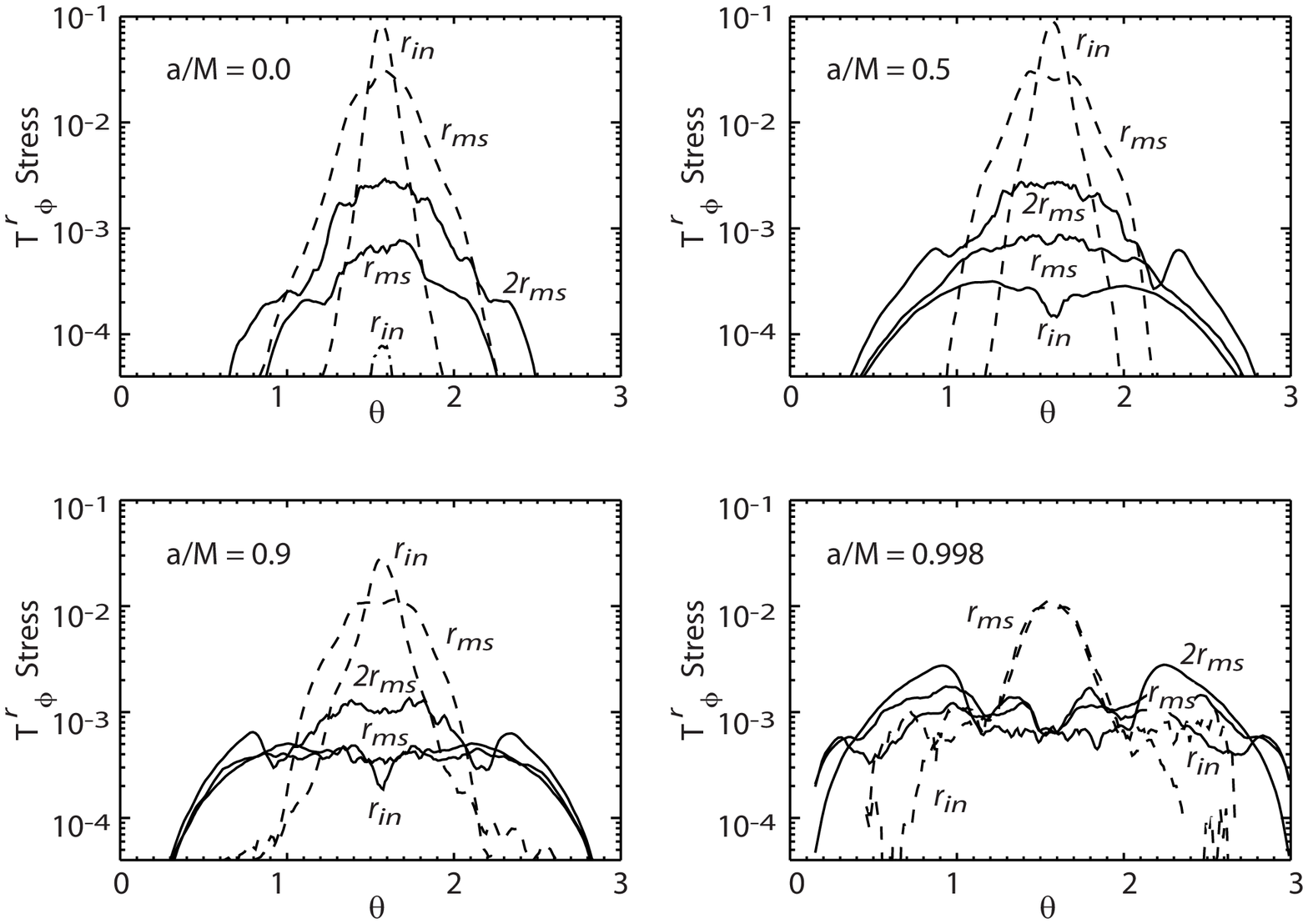}
\caption{\label{blframestress} 
The azimuthally-averaged $T^r_\phi$ stress as a function of polar
angle $\theta$, time-averaged over the last 75\% of the simulation. 
The solid curves are the magnetic contribution to the
outward-directed stress at $r=2r_{ms}$, $r=r_{ms}$ and the inner
boundary.  In the $a/M=0$ model the magnetic stress is inward-directed
(dot-dashed line).  The dashed lines are the mass contribution to the
stress, (i.e., $\rho h U^r U_\phi$), at $r=r_{ms}$ and the inner
boundary.
}
\end{figure}

We can see more precisely where the angular momentum goes by studying
the angular distribution of the magnetic torque contribution to the
stress, $T^r_\phi (r,\theta)$.  
In Figure~\ref{blframestress}
the dashed lines are the (inbound) matter component, plotted at
$2 r_{ms}$ and $r_{ms}$, and the solid lines are the (outbound) total
electromagnetic component, plotted at the same locations as the matter
component and at the inner boundary as well.

The plot for the Schwarzschild model is straightforward.  The matter
angular momentum flux narrows going from $2r_{ms}$ to $r_{ms}$, while
the magnetic stress peaks through the disk and drops off through the
corona.  In the corona there is a region where the outward
electromagnetic stress exceeds the inbound matter flux.  There is
negligible electromagnetic angular momentum flux at the inner boundary,
and what there is is ingoing, represented by a dot-dashed line.

In contrast, all the non-zero spin simulations have a significant
outward angular momentum flux carried by magnetic fields throughout the
region between the marginally stable orbit and the inner boundary.
Moreover, as was also shown in Figure ~\ref{angmomfluxes}, the
magnitude of this flux relative to the stress associated directly with
matter ($\rho h U^r U_\phi$, the dashed curve in the figure) increases
steadily with increasing $a/M$.  As $a/M$ increases, the angular width
through which the magnetic angular momentum flux travels both spreads
and develops a double-maximum structure.  That is, when the black hole
spins slowly, the magnetic angular momentum flux is largely confined to
the region where the matter flux is greatest, a span of perhaps $\pm
30^{\circ}$ on either side of the equatorial plane.  Moving from the
disk to the corona, there is a point where the magnitude of the magnetic
stress exceeds that of the matter stress.  When the black hole spins
rapidly, the magnetic angular momentum flux is relatively constant
with polar angle through the disk region, but rises to a maximum at
the funnel wall and slowly declines in the funnel interior toward the axis.

In the Schwarzschild simulation the magnetic stress peaks within the
disk, and falls as the flow approaches the hole, becoming
inward-directed at the inner boundary.  In the $a/M=0.5$ simulation the
magnetic stress can be described in terms of two pieces: a contribution
spanning a broad angular range that increases slowly with increasing
radius, and another contribution whose angular extent roughly matches
that of the disk matter and grows toward larger radii.  As the black
hole spins faster, the relative importance of the broad angular range
component grows, and a double-peak structure develops, with the
greatest stress occurring near the funnel wall.

We can understand these trends in terms of a changing balance between
effects arising from the magnetorotational instability in the disk and
effects stemming from magnetic coupling to the spin of the black hole
itself.  In the disk, the stress results from MHD turbulence generated
by the MRI (Balbus \& Hawley 1998), and the magnetic stress is tied
directly to the presence of the matter.  The narrow disk component can
be identified with this process; it increases with radius because its
gradient must match that of $\dot M U_\phi$, which increases with
radius.  The broader component we associate with frame-dragging acting
on the magnetic field in the corona and outflow.  Close to the black
hole, the magnetic field intensity varies little with angle, so there
can be stress across a wide range of polar angles.  Its magnitude
changes little with radius because the matter density is so small in
the corona and outflow that there can be little angular momentum
exchange between fields and matter.

In rough terms, it is possible to separate the electromagnetic angular
momentum flux going into the disk from that going into the corona by
evaluating this flux on the shell at $r=r_{ms}$ and distinguishing disk
from corona in terms of polar angle.  Applying this criterion
quantifies some of the remarks of the previous paragraphs: as $a/M$
increases, the ratio of the electromagnetic angular momentum flux
entering the corona to that entering the disk rises from tiny ($<
10^{-3}$ in the Schwarzschild case) to $\simeq 0.3$ ($a/M = 0.5$),
$\simeq 1$ ($a/M = 0.9$) and $\simeq 3$ ($a/M = 0.998$).  Because
angular momentum can readily move from matter to fields and back again,
and because polar angle distribution at a fixed radius does not fully
distinguish the ultimate end of the flux, these numbers are not
entirely well-defined; nonetheless, they do express the qualitative
point that the corona receives a growing proportion as the black hole
spins faster.

When the black hole spins rapidly, the overall level of electromagnetic
angular momentum flux is so high that, despite the growing share
devoted to the corona, the portion going into the disk is large enough
that it fundamentally alters the angular momentum budget of the
accretion flow.  It appears that in both the $a/M = 0.9$ and $a/M =
0.998$ simulations, the time-average magnetic torque {\it adds} to the
matter's angular momentum over a substantial radial range from near the
hole to $r_{ms}$ and beyond (Fig.~\ref{angmomfluxes}).  As a result of
this process, the accretion flow is retarded.  In Paper I, we observed
that in each simulation an inner torus is created in the region
immediately outside $r_{ms}$, and that the prominence of this inner
torus increases with increasing $a/M$.  The electromagnetic stress
driven by the spin of the black hole is the likely origin of this
dependence on $a/M$.  In the case of the $a/M = 0.998$ simulation,
magnetic torques are so strong that the mean of the local angular
momentum can exceed the value associated with a circular orbit.  For
example, the density-weighted ratio of local angular momentum to the
angular momentum of a circular orbit at that radius rises from $\simeq
0.95$ to $\simeq 1.05$ in the inner torus region (see also Fig.~13 of
Paper I).  The effective potential for the flow is then not as flat as
expected in the marginally stable region, and inflow cannot occur as
easily (see also Fig.~\ref{comparetinflow}).

\subsection{Fluid frame stresses}

It is also useful to look at the stresses in the fluid frame.  The
MHD turbulence is primarily due to local instabilities, so its physics
is best analyzed in the local fluid frame.  In addition, 
Novikov \& Thorne (1973) showed that, if the disk
is assumed to be time-steady, axisymmetric, and geometrically thin,
the vertically-integrated stress in the fluid frame can be written in
the physically-enlightening form
\begin{equation}
W^r_\phi (r) = {\dot M \Omega \over 2\pi} R_T(r/M,a/M),
\label{NTstress}
\end{equation}
where the function $R_T$ encapsulates both relativistic corrections
and the net angular momentum flux (the notation here follows Krolik 1999).  
The assumption that stress goes to zero at the marginally stable
orbit means that $R_T(r) = 0$ at all radii $r \leq r_{ms}$.
In the Newtonian limit
of large $r$, $R_T \rightarrow 1$.  In this relation, ``stress" means
the part of the stress tensor other than that directly associated with
accretion, i.e., $W^r_\phi = T^r_\phi - \dot M U_\phi (r)$,
where $U_\phi (r)$ is the circular orbit angular momentum at $r$. 

To compare our results with the Novikov-Thorne prediction, we compute
a closely-related quantity,
\begin{equation}
{\cal W}^{(r)}_{(\phi)} (r) \equiv { \int \, d\theta \, d\phi \,
 \sqrt{-g} \left[
        ||b||^2 U^\mu U_\nu - b^\mu b_\nu\right] 
         \hat e_\mu^{(r)} \hat e^\nu_{(\phi)} 
      \over \int \, d\phi \,
        \sqrt{-g(\theta=\pi/2)/g_{\theta\theta}(\theta=\pi/2)} } ,
\label{sphstress}
\end{equation}
where the $\hat e^\mu_{(\nu)}$ are an orthonormal tetrad designed to
represent the directions labeled by $(\nu)$ in the fluid frame (see
Appendix for details about their construction).  We limit the
integration to the bound portion of the flow, where $-hU_t < 1$.
We divide the shell-integrated
fluid-frame stress by the area per unit radial coordinate in the
equatorial plane in order to give this ${\cal W}^{(r)}_{(\phi)}$ the
same units as $W^r_\phi$.

\begin{figure}[ht]
\centerline{\psfig{file=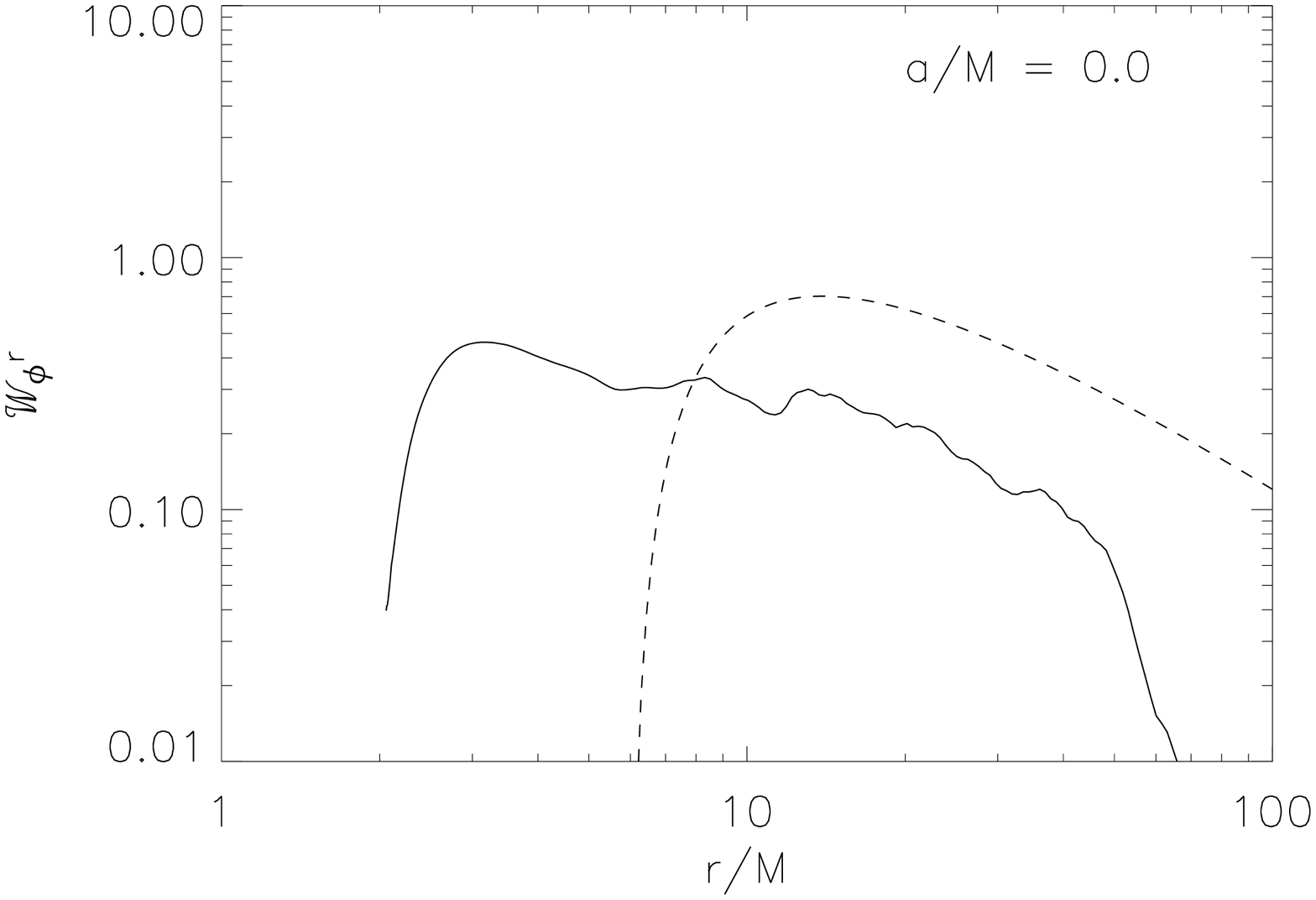,width=2.5in}
       \quad\psfig{file=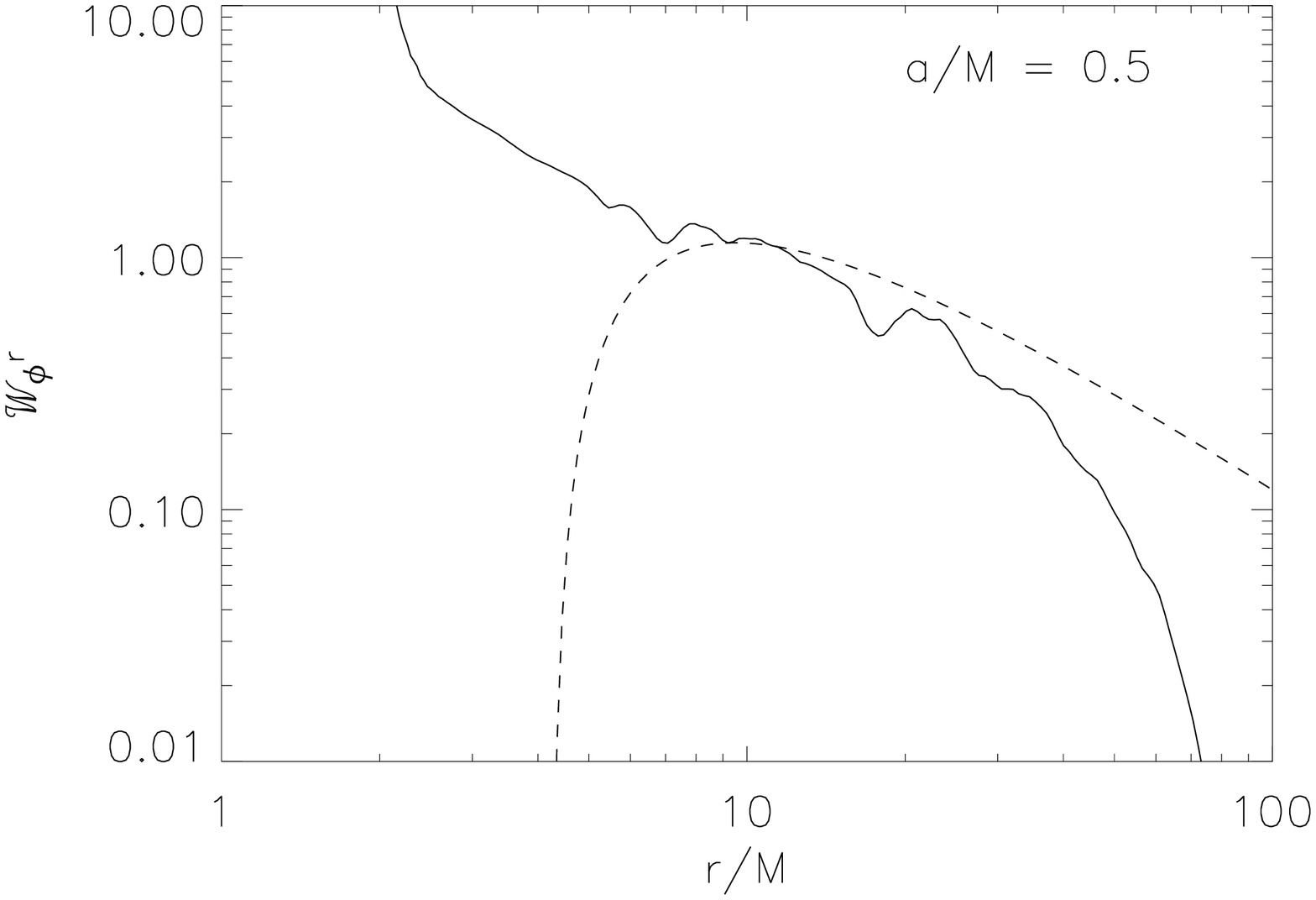,width=2.5in}}
\centerline{\psfig{file=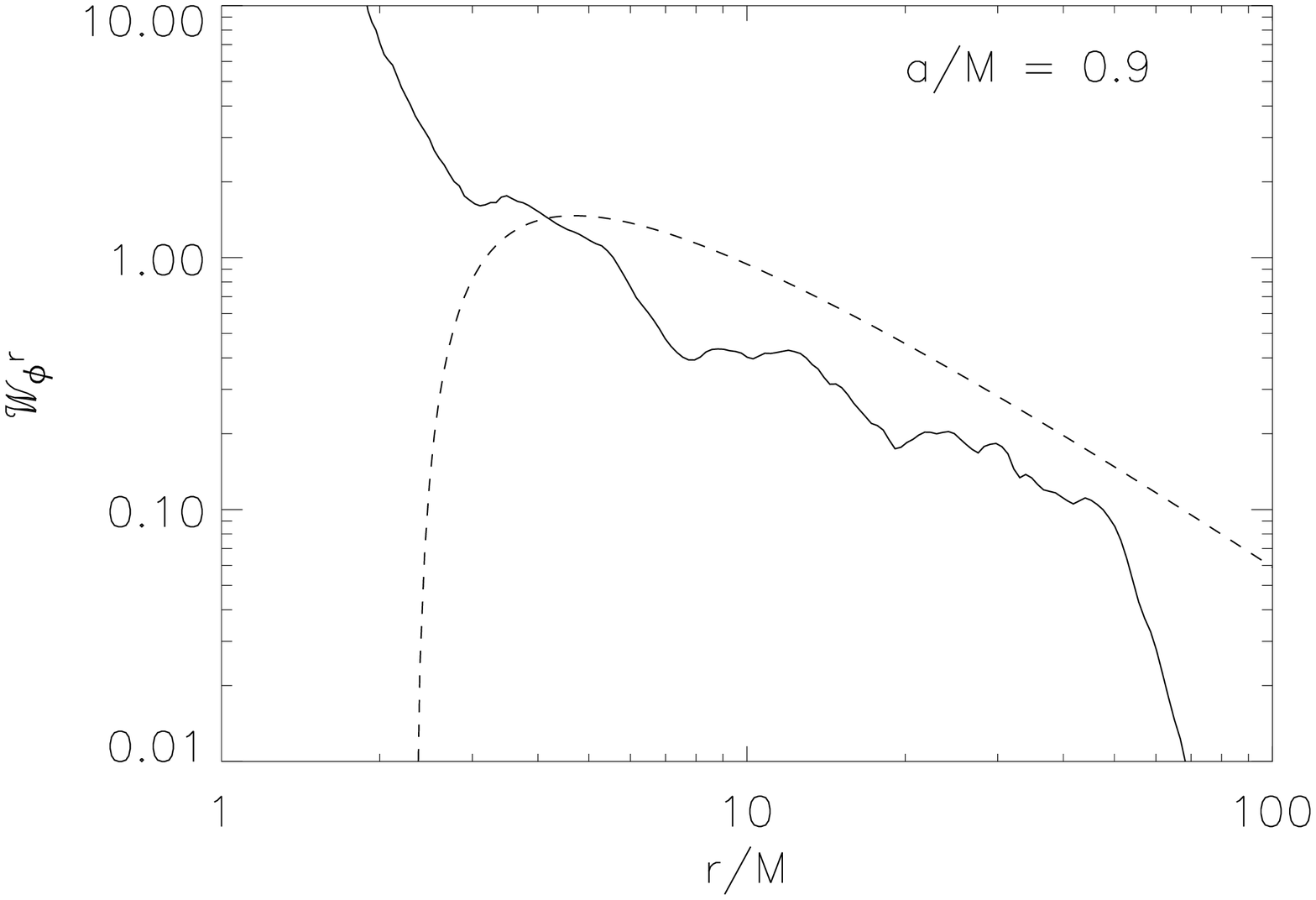,width=2.5in}
       \quad\psfig{file=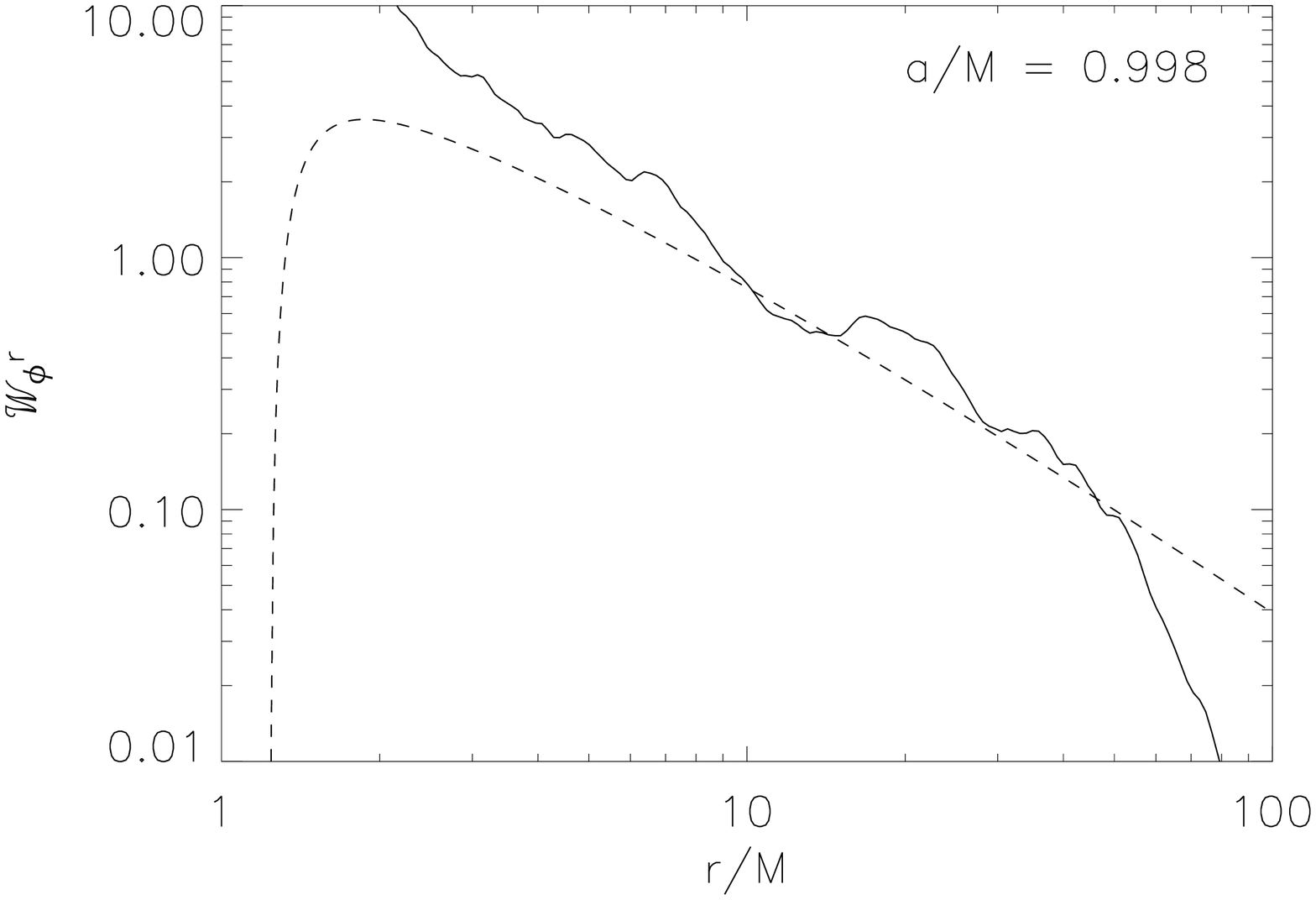,width=2.5in}}
    \caption{\label{flframestress} The shell-integrated magnetic $r$-$\phi$
stress in the fluid frame per unit area in the equatorial plane is plotted
against radius for each of the four simulations at a selected late time.
The dashed curves show the prediction of the Novikov-Thorne model for the
time-average accretion rate at $r_{in}$ for each of these simulations.}
\end{figure}

We compute ${\cal W}^{(r)}_{(\phi)} (r)$ 
at $t = 8000\, M$ for all four
simulations; this is shown in Figure~\ref{flframestress}. 
To facilitate comparison of simulations with different black hole
spins, the value of ${\cal W}^{(r)}_{(\phi)}$ displayed in each panel
is normalized to the initial maximum
pressure found in that simulation.  
There is rough agreement with the Novikov-Thorne prediction (assuming
the time-average accretion rate at the inner boundary) for
$1.5 r_{ms} \lesssim r \lesssim 40M$.  At larger radii, one would not
expect agreement because that is the region in which the matter in our
simulations must move outward to absorb the angular momentum flux,
whereas in the Novikov-Thorne model that region is supplied by
accreting matter from farther away.
Inside $r_{ms}$, however, where the Novikov-Thorne model predicts zero
stress, our data are dramatically different.  We find that the stress
continues quite smoothly, and, in fact, continues to increase 
in amplitude from $r_{ms}$ to the inner boundary
whenever $a/M > 0$.  Only in the Schwarzschild case does the
fluid-frame stress eventually go to zero, albeit 
only very close to the event horizon.

In other words, there is considerable outward angular momentum flux in
the fluid frame virtually throughout the flow, including the plunging
region.  Moreover, whenever $a/M > 0$, there is always at least some
outward angular momentum flux in the fluid frame even at the smallest
radii outside the event horizon.  
Further, even with the overall normalization,
the general level of the fluid-frame magnetic stress
rises substantially with increasing $a/M$.  This is due to the
increasing amplitude of the magnetic fields.
Thus, while there is a fluid-frame stress edge in the
Schwarzschild case just outside the event horizon, {\it
there is no fluid-frame stress edge when the black hole spins}.

The standard Novikov-Thorne model assumes a cold, unmagnetized disk,
with $h = ||b||^2 = 0$.  The origin of the stress is unspecified,
although it is presumed due to turbulent motions.  In our simulations
neither $h$ nor $||b||^2$ is zero.  In the calculation of
Figure~\ref{flframestress} we included only the magnetic stress and
neglected the turbulent Reynolds stress, i.e., $\rho h U^r \delta
U_\phi$, where the perturbation is with respect to a local circular
orbit.  Local simulations (Hawley, Balbus \& Winters 1999) show that
the Reynolds stress is generally a factor of several less than those
due to the magnetic fields.  Obviously the presence of magnetic
fields explains the qualitatively different behavior of ${\cal
W}^{(r)}_{(\phi)}$ and $W^r_\phi$ from the marginally stable region
inward.

Another potential influence on the stress edge is the temperature in
the disk and the associated vertical scale height.  Purely
hydrodynamical models of accretion disks more detailed than the
Novikov-Thorne model (e.g., Muchotrzeb \& Paczy\'nski 1982; Matsumoto
et al. 1984) relax the assumption of $h=0$ and consider the effects of
finite geometric disk thickness.  These studies indicate that, although
the inner boundary for stress may not coincide exactly with the
marginally stable orbit, in thin disks it should lie very close.  Even
in ``slim disks" (Abramowicz et al. 1988), the stress does not extend
all the way across the plunging region: in the pseudo-Newtonian
approximation, designed to mock up Schwarzschild metric dynamics, the
innermost radius for the end of stress stretches inward from $6M$ to
$4M$ as the disk thickens.  But it is clear that the purely
hydrodynamic picture is inadequate;  magnetic stresses operate in ways
that Reynolds stresses and simple viscosity cannot.  Further,
in the MHD simulations there is significant
magnetic field in the surrounding disk corona; the magnetic scale
height is greater than the gas pressure scale height of the disk.  In
fact, the magnetic scale height near the black hole is $\sim r$
more or less independent of the matter scale height in the disk.
Thus, when magnetic fields are included, the picture changes
qualitatively from the purely hydrodynamic predictions.

Earlier we found that, when stress is measured in the coordinate
frame, there is no stress edge anywhere.  When it is evaluated
in the fluid frame, the situation changes only slightly: when
$a/M > 0$, there is similarly no stress edge; when $a/M = 0$,
one exists, but it is deep in the plunging region, immediately outside
the event horizon.

\section{Energy transport}\label{etransport}

\subsection{Energy fluxes}

The conventional approach to defining the radiative efficiency of
accretion onto black holes was set out by Novikov \& Thorne (1973).  In
the standard picture, thermal energy in the disk is assumed negligible
compared to orbital energy, in part because the thermal loss time is
assumed to be very short compared to the inflow time everywhere outside
$r_{ms}$.  With those assumptions, the radiation rate in each ring is
simply the difference between the energy brought into that ring by
mechanical work (torques acting on the ring by neighboring zones) and
energy lost from that ring by accretion.  Both contributions amount, of
course, to the radial divergence of an energy flux which does not
vanish; the energy deposited by the flux is the disk's luminosity.  As
the fluid drops toward the hole, it becomes increasingly bound, with a
binding energy equal to that of the circular orbit at its radius.  The
energy of the last circular orbit at $r_{ms}$ sets the canonical
efficiency of standard thin disk accretion, $\eta =
1+U_t(r_{ms})$.  For a Schwarzschild hole this is $0.057$.

In the simulations, things are less straightforward.  As in the
standard picture, energy is extracted from orbital motion as the gas
accretes, but that energy is not radiated.  It first goes into magnetic
fields and velocity fluctuations within the turbulent disk.  Some
magnetic field rises out of the disk into the corona, and a general
coronal outflow removes some energy from the disk.  As the turbulence
in the disk is dissipated at the grid scale, some of the energy is
converted into heat by artificial viscosity; some is lost.  Enough of
this energy is retained to keep the disk hot.

We can investigate how energy is transported in these simulations by
considering the magnitude of the energy flux, given by the $T^r_t$
component of the stress tensor.  
Just as for the angular momentum flux, it has both a fluid part
\begin{equation}
 \left(T^r_t\right)_{\rm fluid} =  \rho h U^r U_t 
\label{matterenergyflux}
\end{equation}
and an electromagnetic part
\begin{eqnarray}
 \left(T^r_t\right)_{\rm EM} &=& ||b||^2 U^r U_t - b^rb_t \nonumber \\
             &=& F^{\mu r}F_{\mu t} \nonumber \\
             &=&  g^{rr}\left(g^{\phi\phi} {\cal E}^{\phi} {\cal B}^\theta -
              g^{\theta\theta}{\cal E}^\theta {\cal B}^\phi
              - g^{t\phi} {\cal E}^r {\cal E}^\phi\right) ,
\end{eqnarray}
where ${\cal E}^r = V^\phi {\cal B}^\theta - V^\theta {\cal B}^\phi$,
${\cal E}^\theta = V^r {\cal B}^\phi - V^\phi {\cal B}^r$, and
${\cal E}^\phi = V^\theta {\cal B}^r - V^r {\cal B}^\theta$.  The last
form for the electromagnetic energy flux shows that
it is the radial component of the Poynting flux, modified by a
frame-dragging correction.

\begin{figure}[ht]
\epsscale{0.8}
\plotone{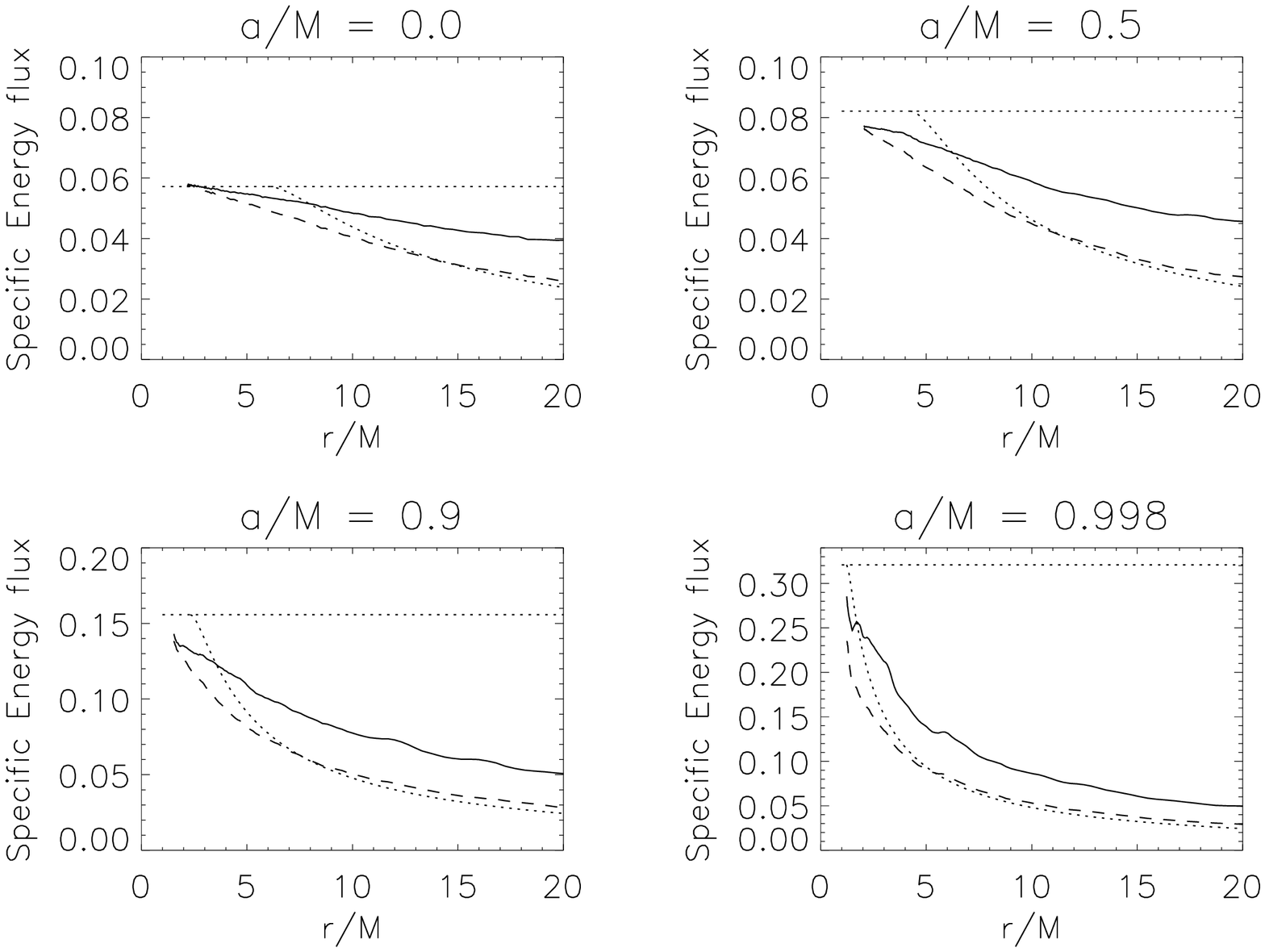}
\caption{\label{specen}
Specific energy flux, defined as $1-\langle\langle T^r_t \rangle\rangle/
\langle\dot M\rangle$ for each 
of the four simulations.  The 
dashed line is derived using the fluid energy flux only, while the
solid line has both the fluid and electromagnetic energy flux.   The
dotted line corresponds to the specific energy of a circular orbit at
each radius; the horizontal dotted line is the specific energy of
the marginally stable circular orbit at at $r_{ms}$.
}
\end{figure}

In a steady-state disk, $T^r_t$ integrated over spherical shells should
be constant with radius.  However, because its largest contribution is
from the flow of rest-mass, its radial behavior is dominated by that
flux, a conserved quantity to which we have already devoted much
attention (\S 3.1).  In order to focus attention on the non-rest-mass
portion, we instead study shell-integrals of time-averaged $T^r_t -
\dot M = \langle\langle \rho(hU_t + 1)U^r + ||b||^2 U^r U_t - b^r
b_t\rangle\rangle$.  In a time-steady disk, the constancy of $\dot M$
with radius means that this ``adjusted" energy flux should likewise be
constant as a function of $r$.  In an effort to make still closer
contact with ordinary intuition, in Figure~\ref{specen} we plot $1 -
\langle\langle T^r_t \rangle\rangle/\langle\dot M\rangle$, the binding
energy per unit rest-mass accreted, but restricting the integral to
cells in which the matter is bound.  In this figure, the dashed line is
the specific energy carried by the fluid term alone
(eq.~\ref{matterenergyflux}), and the solid line is obtained from both the
fluid and electromagnetic components.  As references, there are two
dotted lines: the straight one shows the binding energy of the
marginally stable orbit, the curved one the binding energy of a
circular orbit in the equatorial plane at that radius.

If all the simulations achieved a steady state within the region shown,
and if there were no energy exchange between bound matter and anything
else, the solid line would be flat.  As is immediately apparent, it is
not.  At large radii, the fluid binding energy flux (dashed line)
closely tracks the orbital energy curve.  This is because the numerical
energy loss time is shorter than the inflow time, crudely mimicking the
effects of radiation losses in a real accretion disk.  At smaller
radii, the fluid binding energy is smaller than the orbital binding
energy for two reasons: artificial viscosity adds heat to the gas, and
the inflow time becomes shorter than the numerical loss time.  Some
energy is lost to the outflow, especially when $a/M \geq 0.9$, but not
enough to strongly cool the disk.  Here the mimicry is closer in spirit
to a non-radiative disk.

\subsection{Poynting flux}

\begin{figure}[ht]
\epsscale{0.8}
\psfig{file=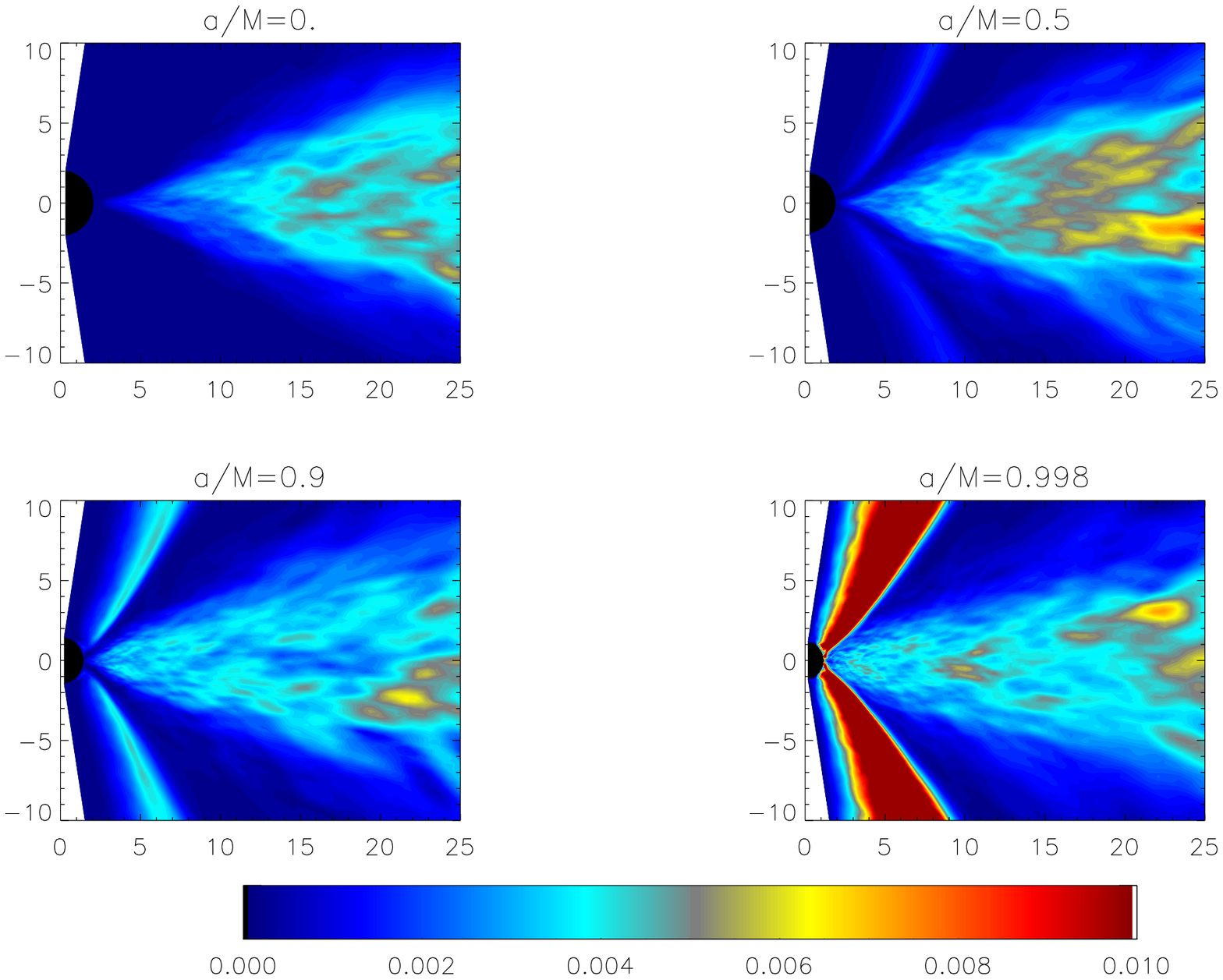,angle=90,width=5.5in}
\caption{\label{poynting}
Time- and azimuthally-averaged outgoing Poynting flux in the four 
simulations, plotted
on the same linear scale.  The values are normalized by the initial
energy and multiplied by $10^4$.
As the black hole spins more rapidly, the Poynting
flux associated with the funnel-wall outflow rapidly grows in importance.
}
\end{figure}

At all radii there is a substantial offset between the
total energy flux curve and the fluid-only curve.  This offset is due to
the outward energy flux conveyed electromagnetically (Fig.~\ref{poynting}).
With the exception of the inner boundary in the zero-spin simulation (where
it is also very small in magnitude), the Poynting flux is always directed
outward.  Independent of the black hole spin, there is always Poynting
flux associated with the work done by torques in the disk.  In
the outer portions of the accreting disk ($r \simeq 15M$), where our
initial conditions make all four simulations very similar, its magnitude
varies little with black hole spin.  However, in the inner disk,
where the mean magnetic field strength grows substantially with
more rapid rotation, so does the Poynting flux.  At the equator on the inner
boundary, it increases by about a factor of two from $a/M = 0.5$ to $a/M =
0.9$ and by almost an order of magnitude from $a/M = 0.9$ to $a/M =
0.998$.

\begin{figure}[ht]
\epsscale{0.5}
\plotone{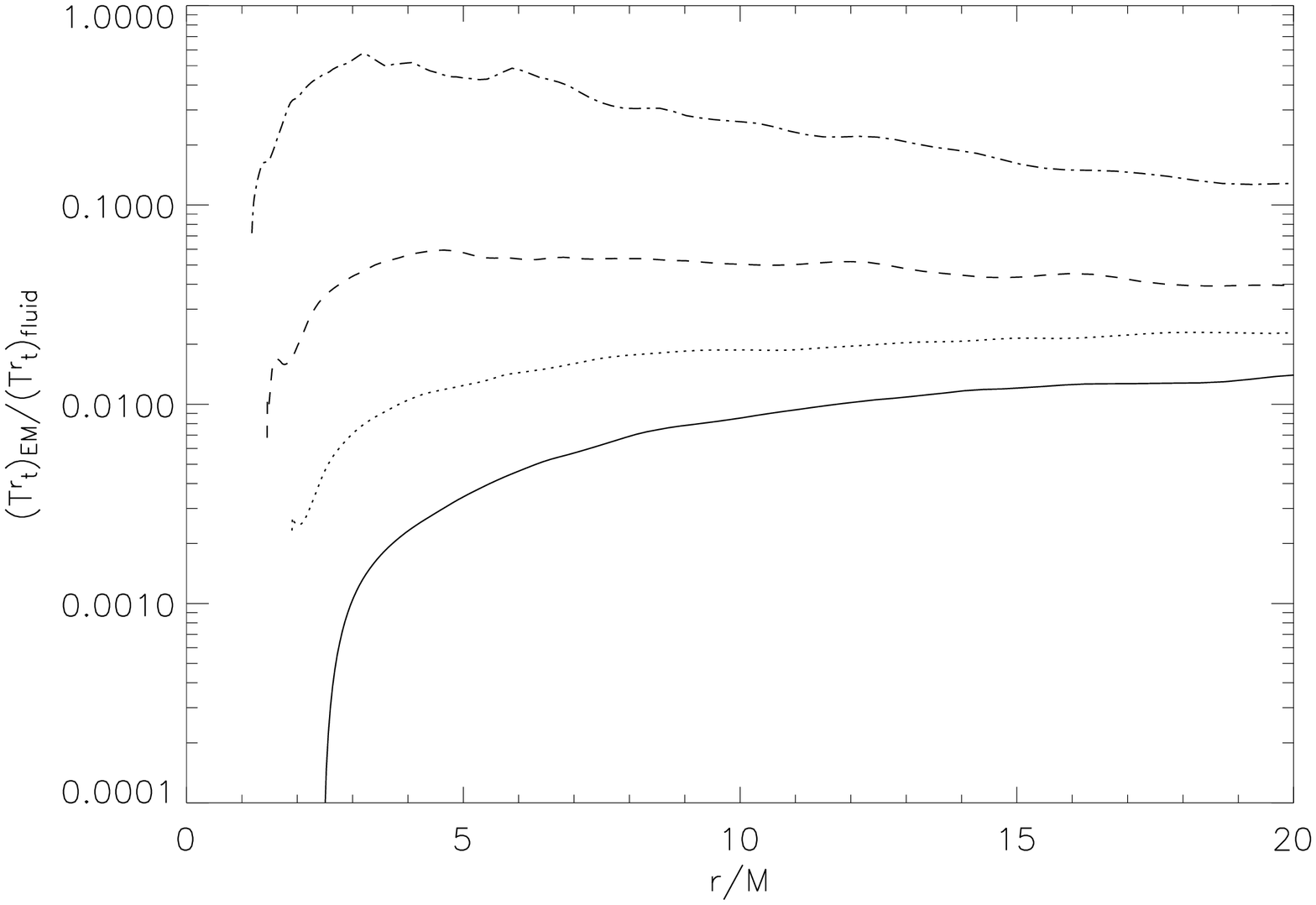}
\caption{\label{etom}
Ratio of time-averaged electromagnetic energy flux, $\langle\langle
T^r_t\rangle\rangle_{EM}$ to fluid energy flux, $\langle\langle
T^r_t\rangle\rangle_{fluid}$, as a function of radius.
The solid line corresponds to the $a/M = 0.0$ run, the dotted line
to $a/M = 0.5$, the dashed line to $a/M = 0.9$ and the  dot-dashed
line to $a/M = 0.998$.  As the spin of the black hole increases, an
increasing fraction of the energy flux is in the form of Poynting
flux.
}
\end{figure}

Poynting flux in the jet outflow varies far more dramatically with
black hole spin.  It is negligible in the Schwarzschild simulation,
while in the $a/M = 0.998$ simulation the Poynting flux in the
funnel wall completely dominates the
Poynting flux in the disk.  Another way to make this comparison
is to consider the ratio of the total (outward) shell-integrated
Poynting flux to the (inward) fluid energy flux (nearly all rest-mass
energy).  This ratio is plotted in Figure~\ref{etom}.
Here we see that relative to the rate at which fluid energy is accreted,
the Poynting flux integrated over the inner boundary rises from 0.25\%
when $a/M=0.5$ to $\sim 1$\% when $a/M=0.9$ to $\sim 10$\% when
$a/M=0.998$.  These ratios increase steadily outward for the $a/M = 0$
and $0.5$ simulations, and have a local maximum around $r=3$--$5\,M$
for the high spin models.  In the simulation with the most rapid spin
at the radius where this ratio reaches its maximum,
Poynting flux carries outward $\sim 60$\% of the ingoing fluid
energy flux.

Another way to look at these results is from the point of view of
the Penrose process.  In its simplest form, this process depends
on the fact that inside ergospheres there can exist negative energy orbits,
so that black holes can give up energy (i.e., lose mass) by absorbing
particles with negative energy.  An analogous process exists for
magnetic fields.  The electromagnetic ``energy density
at infinity" may be defined (Koide 2003) by
\begin{eqnarray}
\label{em_einf}
 e_{\rm EM}^{\infty}/\alpha = - T^t_t (EM) =
                          -&(1/2)g^{tt}\left[ g^{rr} ({\cal E}^r)^2
   + g^{\theta\theta}({\cal E}^\theta)^2 + g^{\phi\phi}({\cal E}^\phi)^2\right]
   + (1/2)\left(g^{t\phi} {\cal E}^\phi\right)^2 \nonumber \\
   &+ \left[ g^{\theta\theta}g^{\phi\phi} ({\cal B}^r)^2
   + g^{rr}g^{\phi\phi}({\cal B}^\theta)^2 + 
      g^{rr}g^{\theta\theta}({\cal B}^\phi)^2\right].\\
\end{eqnarray}
>From this form we see that, if all the spatial elements of the inverse
metric were positive definite, $e_{\rm EM}^{\infty}$ could never be
negative.  However, inside the ergosphere, $g^{\phi\phi} < 0$.  Thus,
negative electromagnetic energy-at-infinity is associated with
regions inside the ergosphere where the components of the field
tensor associated with poloidal magnetic field (${\cal B}^r$ and
${\cal B}^\theta$) and toroidal electric field (${\cal E}^\phi$)
are large in magnitude compared to the other field components.  Moreover,
because ${\cal E}^\phi = V^\theta {\cal B}^r - V^r {\cal B}^\theta$, a large
poloidal magnetic field also helps to make ${\cal E}^\phi$ large, particularly
when the poloidal transport speed is substantial.  That an electromagnetic
version of the Penrose process is at work is suggested by the fact
that in the $a/M = 0.9$ and 0.998 simulations, the time-averaged
$e_{\rm EM}^{\infty} < 0$ in the region
within $40^{\circ}$ -- $60^{\circ}$ of the equatorial plane and $\simeq 0.3M$
of the inner boundary.

One of the key questions of black hole accretion is whether or not 
energy can be extracted from the region inside $r_{ms}$.  Although
it is difficult to say how much radiation will escape from inflowing
gas within the plunging region (discussed below), it is
clear that the Poynting flux provides an avenue for the outward
transport of energy from the plunging region and from the black hole
itself.

\subsection{Radiation}

Although the present simulations do not include radiative losses, it is
interesting nevertheless to consider whether the standard assumptions
of thin disk theory would apply.  The issues involve where and how
rapidly energy is dissipated into heat, and whether there is sufficient
time for the heat to be transformed into photons and radiated prior
to the gas entering the black hole.  Well outside the marginally
stable orbit, the time required for these processes to occur should be
significantly shorter than the inflow time.  Where that is true, the
conservation law approach to computing the radiation rate should work
well.  In the simulations the turbulent energy released by dissipation
is captured only partially by artificial viscosity; numerical losses
thus mimic radiation losses, and in most of the main disk body, the
numerical loss time is appropriately shorter than $t_{in}$.  The only
difference between the simulations and the Novikov-Thorne model in this
regard is that we do not force the stress to go to zero at $r_{ms}$.  A
formalism for tracing the consequences of that changed boundary
condition in time-steady disks was developed in Agol \& Krolik (2000).
Applying it to the simulation results indicates an enhanced radiative
efficiency from this part of the disk that is relatively modest for the
cases with $a/M = 0$, 0.5, and 0.9.  Because the $a/M = 0.998$
simulation did not achieve an inflow steady-state, the assumptions of
this formalism may not be valid, but the data suggest that the
enhancement to the radiative efficiency may be considerably larger.

Inside $\simeq 2r_{ms}$, the inflow time falls rapidly (see
Fig.~\ref{comparetinflow}), becoming roughly an orbital period near and
inside $r_{ms}$.  It then becomes questionable whether the several
processes that stand between inter-ring work and radiation have enough
time to run to completion.  If not, the gas advects its energy inward,
much as envisioned by Rees et al. (1982), Narayan \& Yi (1995), and
Abramowicz et al. (1988), albeit for different reasons.  However,
particularly in the plunging region proper, new processes can lead to
radiation.  Machida \& Matsumoto (2003) suggest that field
reconnection in the plunging region could lead to X-ray flares.
Rather than magnetic field being dissipated on very small
lengthscales at the end of a turbulent cascade, it can instead by
dissipated in large-scale current sheets.  In Hirose et al. (2004) we
drew attention to the prevalence of high current density regions in the
inner disk and plunging region where the current density is so high
that rapid field dissipation might be expected.  If the local
temperature in these current sheets can be raised high enough that
Compton scattering is the dominant cooling mechanism, radiation can
release energy very rapidly:  the Compton cooling time is
\begin{equation} t_C = {3 \over 4} {m_e \over \mu_e}
\left({L \over L_E}\right)^{-1}
	\left({r \over M}\right)^3 M, 
\end{equation} 
where $\mu_e \sim m_p$ is the mass per electron and $L$ is the
luminosity of seed photons injected within radius $r$.  Some of
these photons will be captured by the black hole, and some will strike
the disk farther out, but some portion will also reach distant
observers.  To compute the actual radiative efficiency of
flows of this sort will require calculations that explicitly balance 
the various competing processes and timescales.

\section{Discussion and Summary}\label{summary}

In Paper I of this series we reported that the fraction of the initial
mass accreted in a fixed time diminishes sharply with increasing black
hole spin.  We also noted that the angular momentum per unit rest-mass
that is accreted likewise drops as the black hole spins more rapidly,
and by significantly more than the associated decrease in the specific
angular momentum of the innermost stable circular orbit.  On the basis
of the more detailed analysis presented here, we can now explain both
these effects.  Not surprisingly, they are closely related.

An examination of the angular momentum fluxes in the accretion flow
reveals how significant magnetic fields are to the evolution of the
inner region of a black hole accretion disk.  The low $a/M$ models
attain a quasi-steady state that behaves much as one would expect from
simple accretion disk theory.  The net flux of angular momentum through
the disk is essentially equal to the value carried into the black hole.
Within the flow itself outward-going magnetic angular momentum flux balances
the excess in the inward-going component carried by the matter flow.

Contrary to standard accretion theory, however, there are always
magnetic stresses exerted across the location of the marginally stable
orbit and in the plunging region interior to this point.  In fact, when
the black hole rotates, the $r$--$\phi$ component of the stress tensor
in the fluid frame rises more steeply with decreasing radius in the
plunging region than in the disk itself.  Thus, we have shown that the
concerns voiced quietly in Page \& Thorne (1974) and Thorne (1974) have
been borne out: the presence of magnetic fields invalidates the
``zero-stress" inner boundary condition.

Hirose et al. (2004) showed that the relative magnitude of the magnetic
field in the inner disk grows steadily with increasing black hole
spin.  This is a by-product of the creation of the dense inner torus:
higher pressure gas can support stronger magnetic field when the
field's origin lies in the magneto-rotational instability.  Thus, as
the spin increases, the magnetic stresses grow rapidly stronger.  There
is more going on than can be attributed merely to stronger field
strengths, however.  The rotating hole itself represents a source of
angular momentum and energy that can be carried outward by the magnetic
stress.  When $a/M \gtrsim 0.9$, the magnetic stresses carry out of the
ergosphere an angular momentum flux that is comparable to
what is brought inward with accreting matter.  A striking
suppression of the accretion rate is the result.  In the most extreme case
($a/M = 0.998$), accretion through the inner disk is retarded to such a degree
that even in a simulation time of $8100\, M$ no accretion equilibrium is
achieved; rather, the mass of the inner torus builds steadily through
the simulation as matter accretes from outside, but is held back not
far outside $r_{ms}$.  This result is reminiscent of the suggestion by
van Putten and Ostriker (2001) that magnetic torques driven by rotating
black holes might lead to a state of suspended accretion, but the field
structure we see (turbulent field, largely confined to the accreting
matter) is quite different from the one posited by them (organized
poloidal field with loops spanning all the way from the disk to the
event horizon).

We should note that although the accretion rate over the course of our
simulation is reduced by a factor of 4 contrasting the simulation with
$a/M = 0.998$ to the one with $a/M = 0$, the evident lack of equilibrium
in the former case prevents us
from saying what will happen in the long run.  It is possible that when
the inner torus mass grows large enough, the time-averaged accretion
rate will rise.  In that case, the effect of black hole spin would be
better described as altering the relation between accretion and inner
disk mass rather than as a simple suppression of accretion.  On the
other hand, without doing more extended simulations, we cannot say
whether such an equilibrium is ever achieved---it is possible that the
structure of the inner disk traverses some complicated limit cycle when
the black hole spins very rapidly.  In any case, it is clear that a
magnetized spinning black hole can affect the rate at which matter
accretes into it in ways that go beyond the changes in the effective
potential usually considered.

Another consequence of the substantial magnetic torques near the black
hole is a reduction in the net angular momentum brought to the hole as
a result of accretion.  As mentioned briefly in Paper I and at greater
length in Gammie et al. (2004), the accreted angular momentum per unit
rest-mass decreases so much with increasing $a/M$ that the black hole
can spin down (in the sense of decreasing $a/M$) even as $M$
increases.  Our results are in approximate agreement with those of
Gammie et al. in indicating that there is near balance when $a/M \approx
0.9$, but likely net spin-down for greater $a/M$.  This result suggests
that black holes cannot be spun-up by accretion to $a/M \gtrsim 0.9$;
only a black hole merger might create a more rapidly spinning black
hole.

We also examined energy transport in the vicinity of the black hole.
As with angular momentum transport, the combination of magnetic fields
and rotating holes leads to phenomena not normally included in standard
disk theory.  Magnetic fields can extract both energy and angular
momentum from the black hole in a fashion resembling the
Blandford \& Znajek (1977) mechanism, but with several notable differences.
First, the magnetic field is far from force-free.  Second, the relevant
field lines connect the plunging region to the disk and the outflow
rather than only connecting the event horizon to infinity (cf. similar
results found by McKinney \& Gammie 2004).  Energy is drawn from a
rotating black hole in a manner that, far from being independent of
accretion as in the original Blandford-Znajek picture, is inherently
dependent upon accretion.  In fact, the most convenient way to describe
its magnitude is relative to the rate at which rest-mass is accreted:
for modest spins, the electromagnetic energy release per unit mass is
$\lesssim 1\%$, while for very rapid spin, it can be $\simeq 10\%$ or
more.  As the spin increases, the angular distribution of the Poynting
flux also changes, with most going into the disk at low spin but the
greater share entering the outflow when the black hole spins very rapidly.

The simulations have demonstrated several new dynamical effects for
magnetically driven accretion flows in a Kerr metric.  At this point we
can give only qualitative assessment of the observational consequences
of these effects.  For example, strong outward magnetic angular
momentum flux from a spinning black hole leads to a substantial
alteration in the mass surface density distribution, creating a sizable
inner torus in the region just outside $r_{ms}$.  Completely unexpected
in conventional disk models, this large mass concentration could
significantly alter the expected X-ray reflectivity of this region deep
in the relativistic potential.  Because magnetic stresses continue
throughout the inner disk and plunging region, the efficiency with
which orbital shear energy can be dissipated into heat and then
escaping photons could be greater than conventionally envisioned, but
quantitative results require an improved treatment of thermodynamics.
For rapidly spinning holes, the Poynting flux represents a very
significant part of the energy flow near the black hole.  Although the
impact of these effects on disk thermal emission remains to be worked
out, their presence will surely lead to significant effects not found
in traditional models.

\acknowledgements{We would like to thank Charles Gammie,
Shinji Koide, Masaaki Takahashi, and Ethan Vishniac for 
helpful conversations.  JHK would also like to thank the Institute of
Astronomy, Cambridge for their hospitality while this work was
completed, and the Raymond and Beverly Sackler Fund for support during
his visit there.  This work was supported by NSF grants
AST-0070979 and PHY-0205155, and NASA grant NNG04-GK77G (JFH), and by
NSF grants AST-0205806 and AST-0313031 (JHK and SH). The simulations
were carried out by Jean-Pierre De Villiers on the facilities of the
San Diego Supercomputer Center, supported by the NSF.}

\appendix

\section{Defining the Fluid Frame Tetrads}

    In order to compute quantities measured by a physical observer, one
must define a set of basis vectors (commonly called ``tetrads") appropriate
to that observer, which we write here as $\hat e^{\mu}_{(\nu)}$ for
the contravariant set.  In this notation, $(\nu)$ identifies the direction
in the observer's frame.

    Ideally, these should be directed in four-space so
as to be most useful for physical interpretation.  For example, in most
instances it is important to be able to identify the direction of
proper time in the observer's frame.  This is simply done by setting
$\hat e^\mu_{(t)} = U^\mu$, the observer's four-velocity.  In the observer's
own frame, the only non-zero component of the four-vector is the time
component, so it automatically satisfies the requirement of appropriate
directionality.  Because $U^\mu U_\mu \equiv - 1$, it is also automatically
normalized to unit magnitude.

    Identifying other directions is, in general, more ambiguous.  One can
always construct the remaining three tetrads by a Gram-Schmidt procedure,
requiring the newly-constructed tetrad element to be orthogonal to all
the previously-defined ones and normalized to unit magnitude.  However,
there are many different ways in which the tetrad elements can be ordered
for this procedure and the results depend on the ordering.  In fact,
the results depend in general on whether one first seeks contravariant
or covariant versions.  This mathematical ambiguity emphasizes the physical
ambiguity in identifying spatial directions in one frame with those in another.

    For our purposes, we wish to construct a tetrad basis set that both
reduces to the conventional directions in the Newtonian limit and may be
most directly compared with the set chosen by Novikov \& Thorne (1973).
The choice made by Novikov and Thorne is a special case that is unambiguous
because they were interested only in the limit that $U^r = U^z = 0$ in
the coordinate frame.  When that is so, tetrad elements that are purely
radial or purely vertical in the coordinate frame are automatically orthogonal
to the four-velocity.  Consequently, the $R$ and $z$ tetrad elements
can be taken as purely radial and vertical in the coordinate frame, while
the $t$ and $\phi$ elements couple only to each other.  This so simplifies
the problem that no ambiguities remain.

    Although the components of the fluid four-velocities found in our
simulations are in general non-zero everywhere, in much of the simulation
region, $U^\phi \gg U^r,U^\theta$.  In an effort to ensure that the direction
of greatest difference between the fluid frame and the coordinate frame is
aligned as closely as possible, after setting $\hat e^\mu_{(t)} = U^\mu$,
we begin the Gram-Schmidt process with the $\phi$ element of the tetrad
and set its $r$- and $\theta$-components in the coordinate frame to zero
everywhere.  The $r$-element
can then be constructed with a zero $\theta$-component, but it is also
necessary to insist that $\hat e^r_{(r)} > 0$ to preserve the right
orientation of the tetrad set.  Finally, only the
$\theta$-element remains to be computed; it has non-zero components in
all four coordinates. The result is:

\begin{eqnarray}
e_{(t)}^\mu &=& \left[ U^t, U^r, U^\phi, U^\theta\right] \\
e_{(r)}^\mu &=& (s/N_2)\left[-g^{tt}C_1 - g^{t\phi}C_2,G^{rr}C_0,
            -g^{\phi\phi}C_2 - g^{t\phi}C_1,0\right] \\
e_{(\phi)}^\mu &=& -(1/N_1)\left[l, 0, 1, 0 \right] \\
e_{(\theta)}^\mu &=& (1/N_3)\left[1,(C_1/C_0)(g^{tt} - l^2)/g^{rr},-l,
\right. \nonumber \\
   & & \left.  C_-/C_3 - (U^r/U^\theta)(C_1/C_0)(g^{tt} - l^2)/g^{rr}\right] ,
\end{eqnarray}
where
\begin{eqnarray*}
l &\equiv& U_\phi/U_t  \\
C_0 &\equiv& U^t U_t + U^\phi U_\phi \\
C_1 &\equiv& U^r U_t \\
C_2 &\equiv& U^r U_\phi \\
C_3 &\equiv& U^\theta U_t \\
C_- &\equiv& U^\phi U_\phi - U^t U_t \\
N_1 &\equiv& \left( g_{tt} j^2 - 2 g_{t\phi} j + g_{\phi\phi} \right)^{1/2} \\
N_2 &\equiv& \left(g^{tt} C_1^2 + g^{rr} C_0^2 + g^{\phi\phi} C_2^2 + 
            2 g^{t\phi} C_1 C_2 \right)^{1/2} \\
N_3 &\equiv& \left[g^{tt} - 2g^{t\phi}j  + (C_1/C_0)^2 (g^{tt} - j^2)^2/g^{rr}
     \right.      \\
  & &  + \left.  g^{\phi\phi} j^2 + g^{\theta\theta}(C_-/C_3 - 
              (U^r/U^\theta)(C_1/C_0)(g^{tt} - j^2)/g^{rr})^2\right]^{1/2}\\
s &\equiv& C_0/|C_0|. 
\end{eqnarray*}

\end{document}